\documentclass[12pt]{article}
\usepackage{ssi}
\usepackage{times}
\usepackage{graphicx}

\newcommand{\tltze}{$^{208}$Tl}
\newcommand{\bitof}{$^{214}$Bi}
\newcommand{\dto}{D$_{2}$O}
\newcommand{\hto}{H$_{2}$O}
\newcommand{\nhits}{N$_{hits}$}
\newcommand{\tij}{$\theta_{ij}$}

\newcommand{\nue}{$\nu_{e}$}
\newcommand{\numu}{$\nu_{\mu}$}
\newcommand{\nutau}{$\nu_{\tau}$}
\newcommand{\nux}{$\nu_{x}$}

\newcommand{\phinc}{$\phi_{\mbox{\tiny NC}}$}
\newcommand{\nsix}{$^{16}$N}
\newcommand{\teff}{$T_{\rm eff}$}
\newcommand{\costs}{$\cos\theta_{\odot}$}

\newcommand{\phinumutau}{$\phi_{\mu\tau}$}
\newcommand{\phie}{$\phi_{e}$}
\newcommand{\rtree} {$R^3$}

\newcommand{\snoccfluxshort}{1.76^{+0.06}_{-0.05}\mbox{(stat.)}^{+0.09}_{-0.09}~\mbox{(syst.)}} 
\newcommand{\snoesfluxshort}{2.39^{+0.24}_{-0.23}\mbox{(stat.)}^{+0.12}_{-0.12}~\mbox{(syst.)}} 
\newcommand{\snoncfluxshort}{5.09^{+0.44}_{-0.43}\mbox{(stat.)}^{+0.46}_{-0.43}~\mbox{(syst.)}} 
\newcommand{\snoncfluxunc}{6.42^{+1.57}_{-1.57}\mbox{(stat.)}^{+0.55}_{-0.58}~\mbox{(syst.)}} 
\newcommand{\snomutauflux}{3.41^{+0.45}_{-0.45}\mbox{(stat.)}^{+0.48}_{-0.45}~\mbox{(syst.)}} 
\newcommand{\snoeflux}{1.76^{+0.05}_{-0.05}\mbox{(stat.)}^{+0.09}_{-0.09}~\mbox{(syst.)}} 
\newcommand{\snomutaufluxsk}{3.45^{+0.65}_{-0.62}} 
\newcommand{\snomutaufluxcomb}{3.41^{+0.66}_{-0.64}} 
\newcommand{\nccfit}{1967.7^{+61.9}_{-60.9}} 
\newcommand{\nesfit}{263.6^{+26.4}_{-25.6}} 
\newcommand{\nncfit}{576.5^{+49.5}_{-48.9}}  
\newcommand{\nsigmassno}{5.3} 
\newcommand{\nsigmassk}{5.5} 
\newcommand{\ssmflux}{5.05^{+1.01}_{-0.81}}
\newcommand{\phisk}{2.32\pm0.03\mbox{(stat.)}^{+0.08}_{-0.07}~\mbox{(syst.)}}

\input{epsf.tex}

\begin{document}

\title{Solar Neutrino Observations at the Sudbury Neutrino Observatory}

\author{A.W.P.~Poon\\ 
Institute for Nuclear and Particle Astrophysics \\
Lawrence Berkeley National Laboratory, Berkeley, CA 94720 \\[0.4cm]
Representing the Sudbury Neutrino Observatory Collaboration\footnote{
Q.R.~Ahmad,
R.C.~Allen,
T.C.~Andersen,
J.D.~Anglin,
J.C.~Barton,
E.W.~Beier,
M.~Bercovitch,
J.~Bigu,
S.D.~Biller,
R.A.~Black
% ...
}}

\maketitle

\newpage

\begin{abstract}

The Sudbury Neutrino Observatory (SNO) is a 1000-tonne heavy water Cherenkov detector.  Its usage of \dto\ as target allows the simultaneous measurements of  the \nue\ flux from $^8$B decay in the Sun  and the total flux of all active neutrino species through the charged-current and the neutral-current interactions on the deuterons.  Assuming the standard $^{8}$B shape, the \nue\ component of the $^{8}$B solar neutrino flux is measured to be \phie$=\snoeflux\times 10^6~{\rm cm}^{-2} {\rm s}^{-1}$ for a kinetic energy threshold of 5~MeV.  The non-\nue  component is found to be \phinumutau$=\snomutauflux\times 10^6~{\rm cm}^{-2} {\rm s}^{-1}$.  This $\nsigmassno\sigma$ difference provides  strong evidence for $\nu_{e}$ flavor transformation in the solar neutrino sector.  The total active neutrino flux is measured with the neutral-current reaction at a neutrino energy threshold of 2.2~MeV.  This flux is determined to be \phinc$=\snoncfluxshort\times 10^6~{\rm cm}^{-2} {\rm  s}^{-1}$, and is consistent with solar model predictions.  Assuming an undistorted $^8$B spectrum, the night minus day rate is 14.0$\pm$6.3(stat.)$^{+1.5}_{-1.4}$(sys.)\% of the average rate in the charged-current channel.  If the total active neutrino flux is constrained to have no asymmetry, the night-day asymmetry in the \nue\ flux is found to be 7.0$\pm$4.9(stat.)$^{+1.3}_{-1.2}$(sys.)\%.  A global analysis of all the available solar neutrino data in terms of matter-enhanced oscillations of two active flavors strongly favors the Large Mixing Angle (LMA) solution.

\end{abstract}

\section{Introduction}

For more than 30 years, solar neutrino
experiments~\cite{bib:homestake,bib:kamioka,bib:sage,bib:gallex,bib:gno,bib:superk}
have been observing fewer neutrinos than what are predicted by the
detailed models~\cite{bib:bpb,bib:brun} of the Sun.  This deficit is known as the Solar Neutrino Problem.  A comparison of the predicted and observed solar neutrino fluxes for these experiments
are shown in Table~\ref{tbl:solarnuexp}.  These experiments probe
different parts of the solar neutrino energy spectrum, and show an
energy dependence in the observed solar neutrino flux.  These observations can be
explained if the solar models are incomplete or neutrinos undergo
flavor transformation while in transit to the Earth. 

The Sudbury Neutrino Observatory\cite{bib:sno} was constructed to resolve this 
solar neutrino puzzle.  It is capable of making simultaneous 
measurements of the electron-type neutrino (\nue) flux from $^{8}$B decay 
in the Sun and the flux of all active neutrino flavors through the following reactions:
\[\begin{array}{lcll}
    \nu_{e}+d & \rightarrow & p+p+e^{-} & \hspace{0.5in} \mbox{(CC)}\\ 
    \nu_{x}+d & \rightarrow & p+n+\nu_{x} & \hspace{0.5in} \mbox{(NC)} \\
    \nu_{x}+e^{-} & \rightarrow &  \nu_{x}+e^{-} & \hspace{0.5in} \mbox{(ES)} \\
\end{array}\]
The charged-current (CC) reaction on the deuteron is sensitive exclusively to \nue, 
and the neutral-current (NC) reaction has equal sensitivity to all 
active neutrino flavors (\nux ; $x=e,\mu,\tau$).  Elastic scattering 
(ES) on electron is also sensitive to all active flavors, but with 
reduced sensitivity to \numu\ and \nutau.  

\begin{figure}
    \begin{center}
    \includegraphics[width=5.75in]{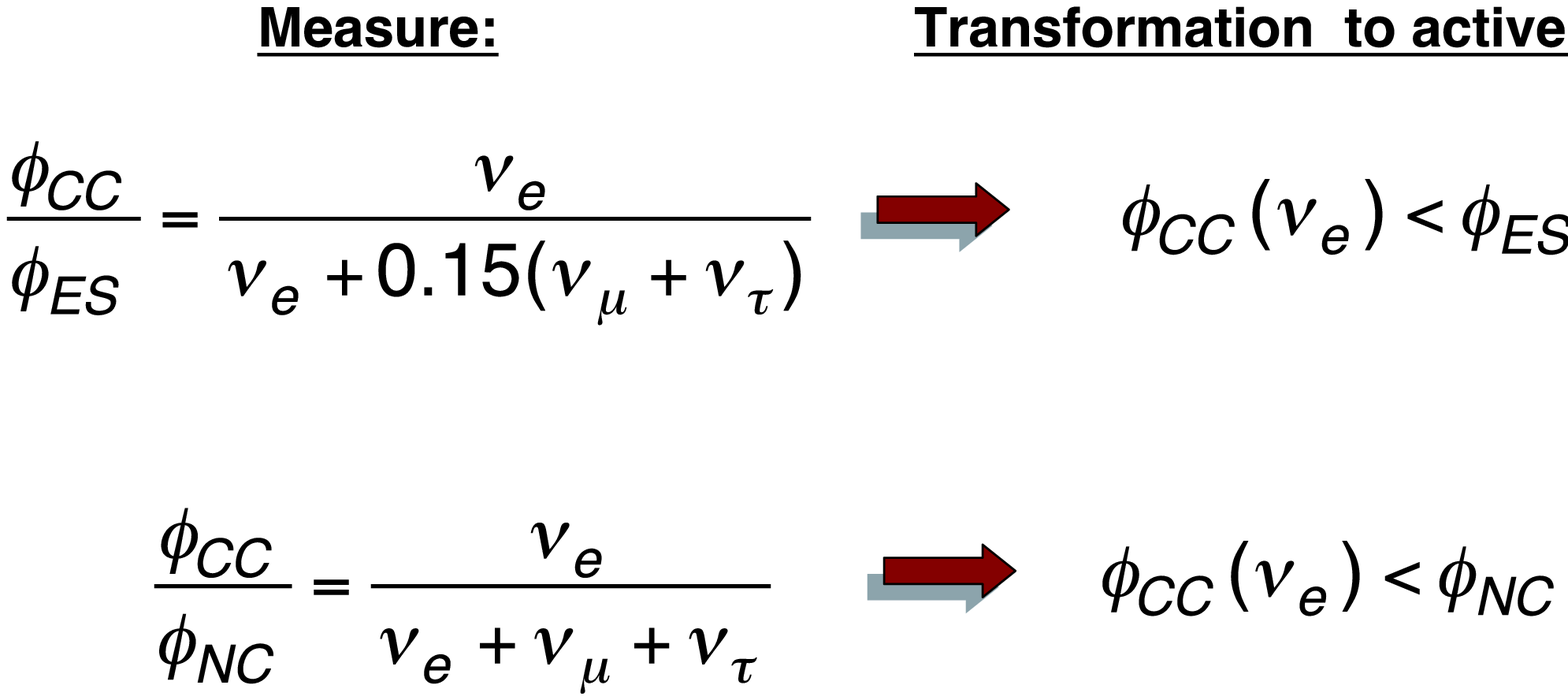}    
   \end{center}
    \caption{Using the measured solar neutrino fluxes from different 
    reaction channels to provide ``smoking gun'' evidence of neutrino 
    flavor transformation.}
    \protect\label{fig:smoking_gun}
\end{figure}

Because of its equal sensitivity to all active neutrinos, the neutral-current measurement can determine the total neutrino flux, hence resolving the Solar Neutrino Problem, {\it even if neutrinos oscillate}\cite{bib:chenprl}.  SNO is currently the only experiment that can simultaneously observe the {\it disappearance} of \nue\ and the {\it appearance} of another active species.  This is illustrated in Figure~\ref{fig:smoking_gun}.    Another feature of the neutral-current interaction is its low kinematic threshold.  By efficiently counting the free neutrons in the final state of the neutral-current reaction, the total active $^8$B neutrino flux can be inferred for neutrinos with energy above the 2.2-MeV kinematic threshold.

Recent results~\cite{bib:snocc,bib:snonc} from the measurements of the solar $^{8}$B neutrino flux by the SNO detector using the CC, NC and ES reactions are presented in this 
paper.  The results\cite{bib:snodn} of a measurement of the day-night asymmetry of the neutrino event rates, which is predicted under certain neutrino oscillation scenarios, are also presented.  Finally, the physics implications of these observations will be discussed.

\begin{table}[tp]
    \begin{center}
    \begin{tabular}{llll} \hline
	Experiment & Measured Flux & SSM Flux~\cite{bib:bpb}\\ \hline
	Homestake\cite{bib:homestake}  &  
	2.56$\pm$0.16(stat.)$\pm$0.16(sys.) SNU & 7.6$^{+1.3}_{-1.1}$SNU
	 \\ \hline
	SAGE\cite{bib:sage} & 70.8$^{+5.3}_{-5.2}$(stat.)$^{+3.7}_{-3.2}$ SNU & 128$^{+9}_{-7}$ SNU  \\ \hline
	Gallex\cite{bib:gallex}  & 77.5$\pm$6.2(stat.)$^{+4.3}_{-4.7}$(sys.) SNU & 128$^{+9}_{-7}$ SNU \\ 
	GNO\cite{bib:gno} & 65.8$^{+10.2}_{-9.6}$(stat.)$^{+3.4}_{-3.6}$(sys.) SNU & 128$^{+9}_{-7}$ SNU  \\ \hline
	Kamiokande\cite{bib:kamioka}  & 	2.80$\pm$0.19(stat.)$\pm$0.33(sys.)$\times$10$^{6}$ & 
	5.05$\times 10^{6}\left(1^{+0.20}_{-0.16}\right)$ \\ 
	Super-Kamiokande\cite{bib:superk} & 	
	2.35$\pm$0.03(stat.)$^{+0.07}_{-0.06}$(sys.)$\times$10$^{6}$ & 
	5.05$\times 10^{6}\left(1^{+0.20}_{-0.16}\right)$ \\ \hline
    \end{tabular}
   \end{center}
    \caption{Summary of solar neutrino observations at different 
    solar neutrino detectors.  The measured fluxes at the radiochemical experiments are measured in Solar Neutrino Unit (SNU), which is defined as 1 capture per second per 10$^{36}$ target atoms.  The Kamiokande and Super-Kamiokande measurements are in units of 10$^{6}$~cm$^{-2}$~s$^{-1}$.}
    \protect\label{tbl:solarnuexp} 
\end{table}

\section{The Sudbury Neutrino Observatory}

\subsection{Physical Description of the SNO Detector}

SNO~\cite{bib:sno} is an imaging water Cherenkov detector located in
the Creighton \#9 mine, owned by the International Nickel Company
(INCO) near Sudbury, ON, Canada.  A barrel-shaped cavity with a height
of 34~m and a diameter of 22~m was excavated at a depth of 2092~m (or
6000 meters of water equivalent) to house the detector.  The muon flux
traversing the detector is 67~day$^{-1}$.

\begin{figure}
    \begin{center}
    \includegraphics[width=5.75in]{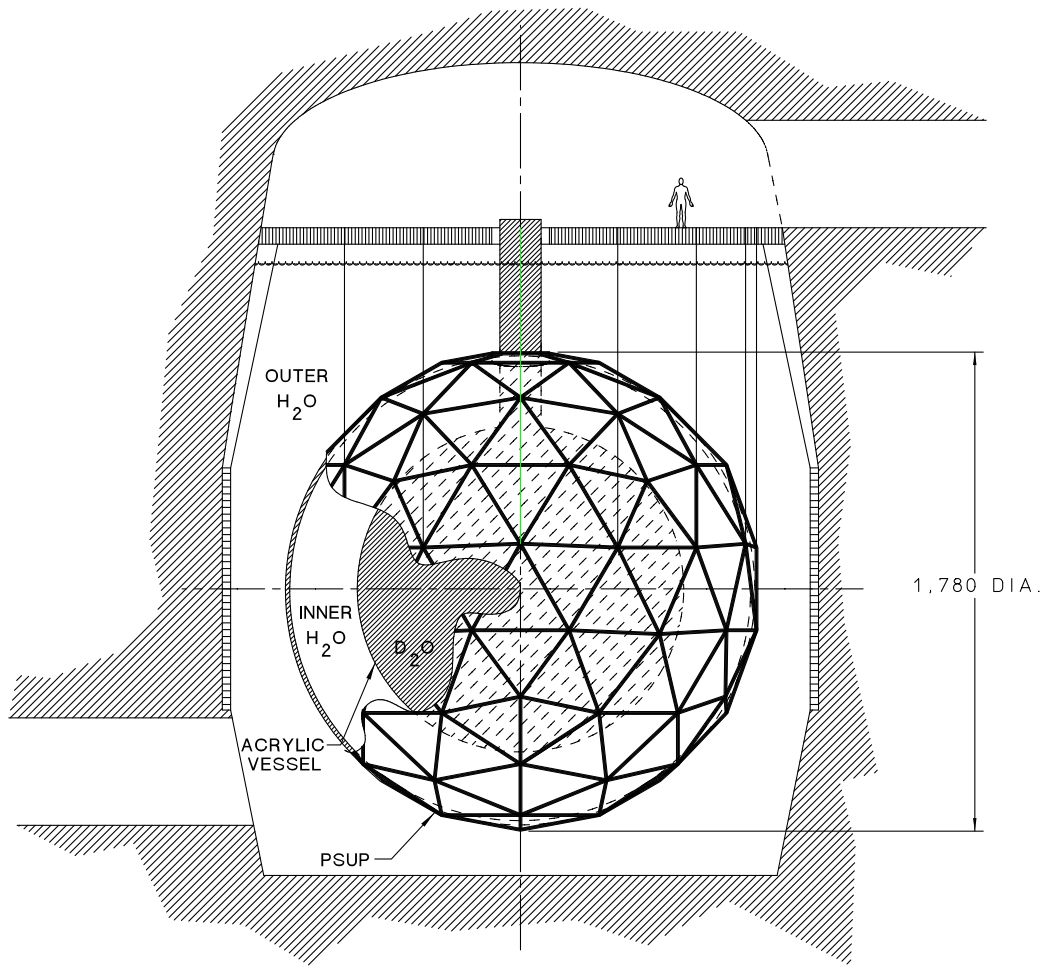}    
    \end{center}
    \caption{A cross-sectional view of the SNO detector.  The outer 
    geodesic structure is the PMT support (``PSUP'').}
    \protect\label{fig:SNO-SK0126D_cm}
\end{figure}

Figure~\ref{fig:SNO-SK0126D_cm} shows a cross-sectional view of the
SNO detector.  The neutrino detection medium is 1000 metric tons of
99.92\% isotopically pure \dto\ contained in a 12-m diameter acrylic
sphere.  The acrylic vessel is constructed out of 122 ultraviolet
transmitting acrylic panels.  This sphere is surrounded by 7000 metric
tons of ultra-pure \hto\ contained in the cavity.  This volume of
\hto\ shields the detector from high energy $\gamma$ rays and neutrons
originating from the cavity wall.  A 17.8-m diameter stainless steel
structure supports 9456 20-cm diameter inward-facing photomultiplier tubes
(PMTs).  A non-imaging light concentrator is mounted on each PMT, extending the total photocathode coverage to 55\%.  An additional 91 PMTs are
mounted facing outward on the support structure to serve as a cosmic-ray 
veto.  To cancel the vertical components of the terrestrial magnetic
field, 14 horizontal magnetic compensation coils were built into the
cavity wall.  The maximum residual field at the PMT array is
$<$19$\mu$T, and the reduction in photo-detection efficiency is about
2.5\% from the zero-field value.

A physics event trigger is generated in the detector when there are 18 or more
PMTs exceeding a threshold of $\sim$0.25 photo-electrons within a
coincidence time window of 93~ns.  All the PMT hits registered in the
$\sim$420~ns window after the start of the coincidence time window are 
recorded in the data stream.  This widened time window allows 
scattered and reflected Cherenkov photons to be included in the event.  The mean noise rate 
of the PMTs is $\sim$500~Hz, which results in $\sim$2 noise PMT hits in 
this 420~ns window.  The instantaneous trigger rate is about 
13-20~Hz, of which 5-8~Hz are physics triggers.  The remaining 
triggers are diagnostic triggers for monitoring the well 
being of the detector.  The trigger efficiency reaches 100\% when the PMT
multiplicity (\nhits) in the event window is $\geq$23.

\subsection{Solar Neutrino Physics Program at SNO}

The solar neutrino physics program at SNO is designed to exploit its 
unique NC capability.  Because the result of this NC 
measurement is a definitive statement on the Solar Neutrino Problem and flavor transformation of solar neutrinos, the SNO experiment has implemented a plan to make three independent NC measurements of the total $^{8}$B active neutrino flux.

The first NC measurement was made with a pure \dto\ target.  The 
free neutron from the NC interaction is thermalized, and for  
$\sim$30\% of the time, a 6.25-MeV $\gamma$ ray is emitted following 
the neutron capture by a deuteron.  Only $\sim$50\% of the 
6.25-MeV observed photopeak is above the neutrino analysis threshold, yielding a detection efficiency of $\sim$15\%.  The results from this NC measurement is presented in this paper.  The 
second NC measurement is being made with 2 tonnes of NaCl added to the \dto.  The free 
neutron is readily captured by $^{35}$Cl in this detector configuration, and a cascade 
of $\gamma$ rays with a total energy of 8.6~MeV follow.  The neutron 
detection efficiency is significantly enhanced, and $\sim$45\% of the 
NC events have a detectable signal above the analysis threshold.  This phase of the experiment is scheduled to complete by the end of 2002.   In the third NC measurement, discrete $^{3}$He proportional counters will be installed inside the \dto\ volume~\cite{bib:ncd}.  The neutron detection efficiency of the proportional counter array is $\sim$40\%.  In this detector configuration, 
the detection of the CC and the NC signals are decoupled, and the 
covariance of the CC and NC signals that appear in the first two 
detector configurations is eliminated in this case.

\section{Solar Neutrino Analysis at SNO}

\begin{figure}
    \begin{center}
    \includegraphics[width=5in]{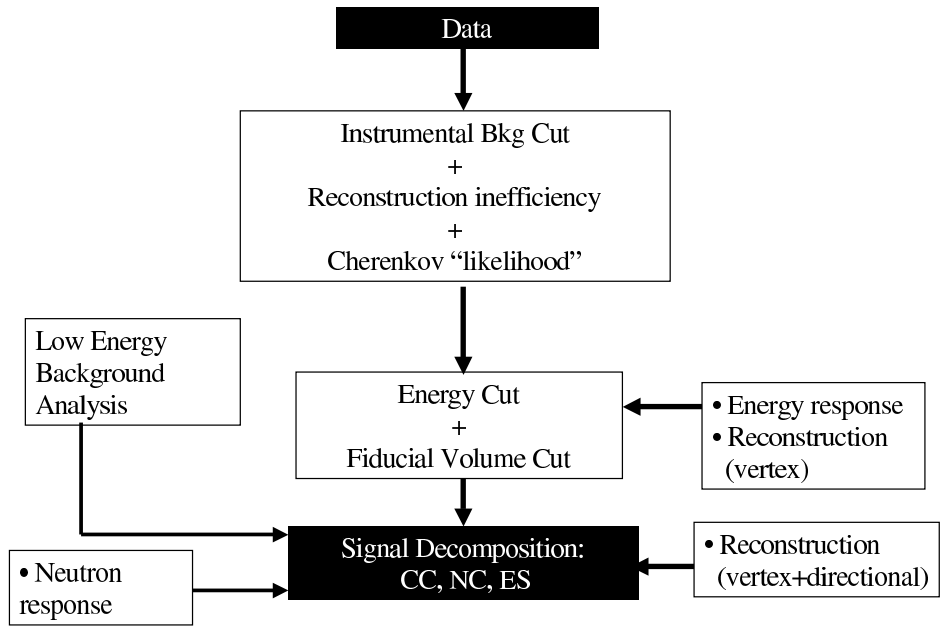}    
    \end{center}
    \caption{Simplified flow chart of solar neutrino analysis at SNO.}
    \protect\label{fig:analysis_flow}
\end{figure}

The data presented in this paper were recorded between November 2,
1999 and May 28, 2001.   The corresponding livetime is 
306.4~days.  During this data acquisition period, the Sun was above the detector's horizon (``day" data set) for 128.5~days, and below the detector horizon (``night" data set) for 177.9~days.   The target was pure \dto\ during this period.  Figure~\ref{fig:analysis_flow} summarizes the analysis procedure for extracting the CC, NC and the ES event rates in the SNO detector.  In the following, each step in the analysis flow is discussed in detail.

\subsection{Data Reduction and Data Loss}

After removing all the detector diagnostic triggers from the data
stream, a significant portion of the remaining events are instrumental
backgrounds.  Examples of these backgrounds include electrical discharges in the PMTs (``flashers'') and 
light emission from discharges in insulating detector materials.  Data reduction schemes were developed to remove these backgrounds.  

\subsubsection{Pass 0 Cut}

The instrumental backgrounds have
characteristic PMT time and charge distributions that are
significantly different from Cherenkov light, and can be eliminated
using cuts based on these distributions.  For example, the discharge
light emitted from a flasher PMT is detected across 
the detector $\sim$70~ns after the initial discharge is registered. 
Some of these light-emitting instrumental backgrounds are localized
near the water piping near the top of the detector.  Veto PMTs were
installed in this region in order to enhance the rejection efficiency
of these non-Cherenkov events.  Interference in the electronics system
can produce false events.  Most of the registered electronic channel
charges in these interference events are near the pedestal, and can be
removed by a cut on the mean charge of the fired PMTs.  Some of these 
electrical discharge or electronic interference background events also 
have different event-to-event time correlations from physics events, 
and time correlation cuts are used to remove these events.  Two 
independent instrumental background rejection schemes are used.  An 
event-by-event comparison of the data sets reduced by these two 
schemes shows a difference of $<$0.2\%.  

The physics loss due to these instrumental background cuts is
calibrated with a triggered \nsix\ 6.13-MeV $\gamma$-ray 
source~\cite{bib:nsix} and a
triggered $^{8}$Li 13-MeV endpoint $\beta$ source~\cite{bib:tagg} deployed to the
\dto\ and \hto\ volumes.  Further tests of the \nhits\ dependence in 
the cuts are performed with an isotropic light source at various 
intensities.  The physics acceptance of the instrumental
background cuts, weighted over the fiducial volume, is measured to be 0.9966$^{+0.0011}_{-0.0002}$.  
Instrumental background rejection is well over 99\% at this stage.

\subsubsection{High Level Cuts}

After passing the instrumental background cuts, all events with
\nhits$\geq$30 ($\sim$3.5~MeV electron energy) are reconstructed.  
Once the event reconstruction information becomes available after the
reconstruction, several high level physics cuts are applied to the
Pass 0-reduced data set to further reduce the instrumental 
backgrounds.  The efficiency of the reconstruction algorithm is calibrated with the \nsix\ and the $^8$Li sources, and is found to be 0.9985$\pm$0.0015 for neutrino events originated from the central 550~cm of the detector.

The high level cuts test the hypothesis that each 
event has the properties of electron Cherenkov light.  
The reconstruction figure-of-merit cuts test for the consistency
between the time and angular expectations for an event fitted to the 
location of the reconstructed vertex and that based on the properties of
Cherenkov light and the detector response.  

\begin{figure}
    \begin{center}
     \includegraphics[width=3.25in]{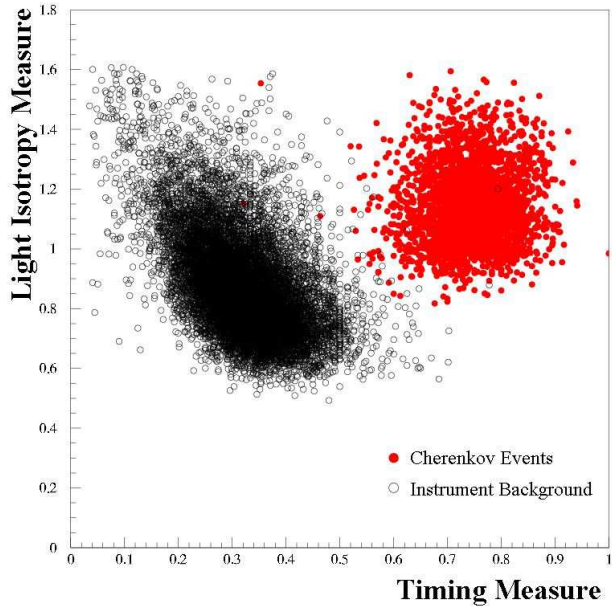}    
    \end{center}
    \caption{Separation of instrumental backgrounds and Cherenkov 
    light events using the high level cuts.}
    \protect\label{fig:hilcuts2}
\end{figure}

Two parameters are used to further characterize Cherenkov light.  The
average opening angle between two hit PMTs ($\langle$\tij$\rangle$),
measured from the reconstructed vertex, is used to determine whether
the topology of an event is consistent with Cherenkov light.  The
in-time ratio (ITR) is the ratio of the number of hit PMTs within an
asymmetric time window around the prompt light peak to the number of
calibrated PMTs in the event.  Figure~\ref{fig:hilcuts2} shows the
correlations between \tij\ and ITR for instrumental backgrounds and
Cherenkov light events.  As shown in the figure, this two dimensional
cut has very high instrumental background rejection efficiency.

The total signal loss from the Pass 0 and the high level cuts are
calibrated with the $^{16}$N and the $^{8}$Li sources.  Because of the difference in the energy spectrum and spatial distribution for CC, ES and NC events, there is a slight difference in the data loss due to the Cherenkov likelihood cut for the different detection channels.  Table~\ref{tbl:phy_accept} summarizes the physics acceptance after the Pass-0, reconstruction and high level cuts for the three neutrino detection channels.

\begin{table}[tp]
\begin{center}
\begin{tabular}{l|l|l|l|l}\hline
Channel  &  Pass-0   &  Reconstruction  &  High Level Cuts &  Total \\ \hline
CC & 0.9966$^{+0.0011}_{-0.0002}$ & 0.9985$\pm$0.0015 & 0.9906$\pm$0.0005 & 0.9857$^{0.0039}_{0.0021}$ \\ \hline
ES & 0.9966$^{+0.0011}_{-0.0002}$ & 0.9985$\pm$0.0015 & 0.9903$\pm$0.0007 & 0.9854$^{0.0040}_{0.0023}$ \\ \hline
NC & 0.9966$^{+0.0011}_{-0.0002}$ & 0.9985$\pm$0.0015 & 0.9821$\pm$0.0011 & 0.9772$^{0.0041}_{0.0023}$ \\ \hline
\end{tabular}
    \end{center}

\caption{Physics acceptance at each of the instrumental background removal stages.  A kinetic threshold of 5~MeV and a fiducial volume of the inner 550~cm of the detector are assumed.}
\protect\label{tbl:phy_accept}

\end{table}

The residual instrumental background contamination in the neutrino
signal after the series of instrument background cuts is estimated by a
bifurcated analysis, in which the signal contamination is obtained
from cross calibrating the background leakage of two groups of
orthogonal cuts.  For the same fiducial volume and energy threshold,
the instrumental background contamination is estimated to be
$<$3~events (95\% C.L.), or 0.1\% of the final neutrino candidate data
set.  Table~\ref{tbl:data_reduction} summarizes the sequence of cuts 
that are used to reduce the raw data to 2928 neutrino candidate 
events.

\begin{table}
    \begin{center}
    \begin{tabular}{lr} \hline
	{\bf Analysis step} & {\bf Number of events} \\ \hline
	Total event triggers & 450 188 649 \\
	Neutrino data triggers & 191 312 560 \\
	PMT hit multiplicity (\nhits) $\geq$30 cut & 10 088 842\\
	Instrumental background (Pass 0) cuts & 7 805 238 \\
          Cherenkov likelihood cuts &  3 418 439 \\
	Fiducial volume cut & 67 343 \\
	Energy threshold cut & 3 440 \\
	Cosmic-induced background subtraction & 2 928  \\ \hline
	{\bf Total events} & {\bf 2 928} \\ \hline
    \end{tabular}
   \end{center}
    \caption{Number of candidate events remained after each data reduction step}
    \protect\label{tbl:data_reduction}
\end{table}

\subsection{Reconstruction and Energy Calibration}

After all the instrumental background cuts have been applied to the data, energy and fiducial volume cuts are employed to reduce the  physics backgrounds in the neutrino candidate data set.  These physics backgrounds include low energy backgrounds from radioactive decays in the natural $^{238}$U and $^{232}$Th  chains, and other backgrounds induced by cosmic rays.  Because of progressively lower radioactivity towards the central region of the detector, one can place a fiducial volume cut and an energy threshold sufficiently high that most of the low energy backgrounds are removed from the neutrino candidate data set.  These cuts rely upon event reconstruction and detector energy calibration, which we will discuss in the following.

\subsubsection{Reconstruction}

The calibrated times and positions of the fired PMTs are used to
reconstruct the vertex position and the direction of the particle. 
Two different reconstruction algorithms were developed.  An
event-by-event comparison shows excellent agreement between the data
sets reconstructed by these two algorithms.  The neutrino data presented in
this paper are reconstructed using a maximum likelihood technique
which uses both the time and angular characteristics of Cherenkov
light.  Vertex reconstruction accuracy and resolution for electrons
are measured using Compton electrons from the \nsix\ $\gamma$-ray
source, and their energy dependence is verified by the $^{8}$Li
$\beta$ source.  Compton scattered electrons from a 6.13-MeV $\gamma$
ray are preferentially scattered in the forward direction relative to
the incident $\gamma$-ray direction.  Thus they provide information about the accuracy of the direction reconstruction.  In order to minimize the effect
of finite vertex resolution on this angular resolution measurement,
only $^{16}$N events that are reconstructed to more than 150~cm from
the source are used in the measurement.  At the $^{16}$N energy
($\sim$5.5~MeV total electron energy), the vertex reconstruction
resolution is 16~cm and the angular resolution is 26.7$^{\circ}$.  
Reconstruction-related systematic uncertainties to the solar neutrino 
flux measurement are $^{+2.9}_{-2.8}$\% for the CC channel and  $^{+1.8}_{-1.8}$\% for the NC channel.

\subsubsection{Energy Estimator}

\begin{figure}
    \begin{center}
     \includegraphics[width=4.5in]{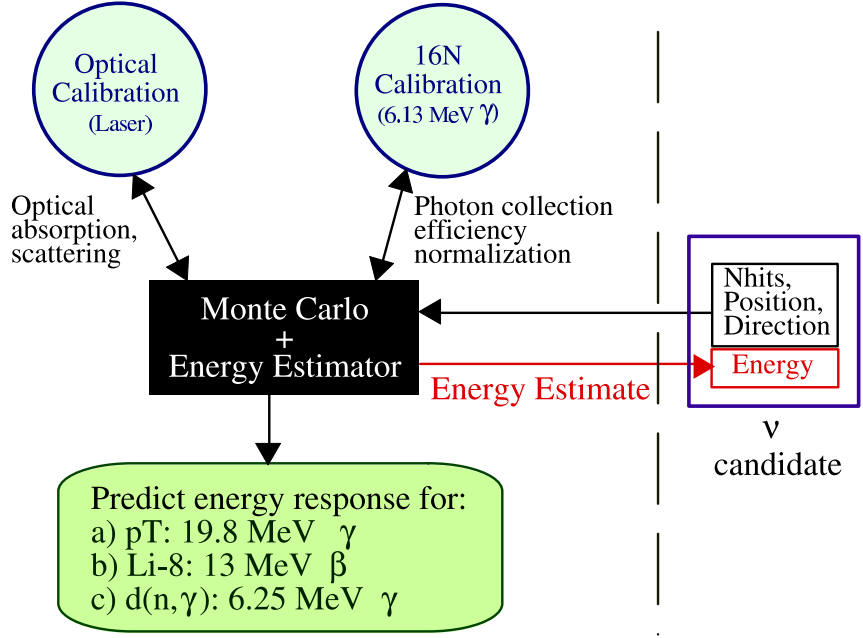}    
    \end{center}
    \caption{Calibration of the SNO detector and event-by-event 
    energy estimator.}
    \protect\label{fig:calib_flow_chart}
\end{figure}

Figure~\ref{fig:calib_flow_chart} shows the relationship between the
detector calibration program and event-by-event energy estimation in
the analysis.  Optical calibration is obtained using a near-isotropic
source of pulsed laser light~\cite{bib:ford,bib:fordphd} at 337, 365,
386, 420, 500 and 620~nm.  The light source is deployed to locations
accessible by the source manipulator system on two orthogonal planes in
the \dto, and on a linear grid in the \hto.  Optical parameters of 
different optical media in the detector are obtained at these
wavelengths~\cite{bib:moffat}.  The attenuation lengths in \dto\ and
\hto\ are found to be near the Rayleigh scattering limit.  These
optical parameters are inputs to the Monte Carlo/energy estimator engine.

The triggered \nsix\ source  is used to
provide the absolute energy calibration.  The detector energy response
to the photopeak of this source provides a normalization to the PMT
photon collection efficiency used in the Monte Carlo, and establish
the absolute energy calibration.  A long-term stability study of the
detector response to the \nsix\ source shows a linear drift of
1.3\%~year$^{-1}$.  The cause of this effect is under
investigation, and a drift correction is applied to the event-by-event
energy estimator.

\begin{figure}
    \begin{center}
 \includegraphics[height=3.5in]{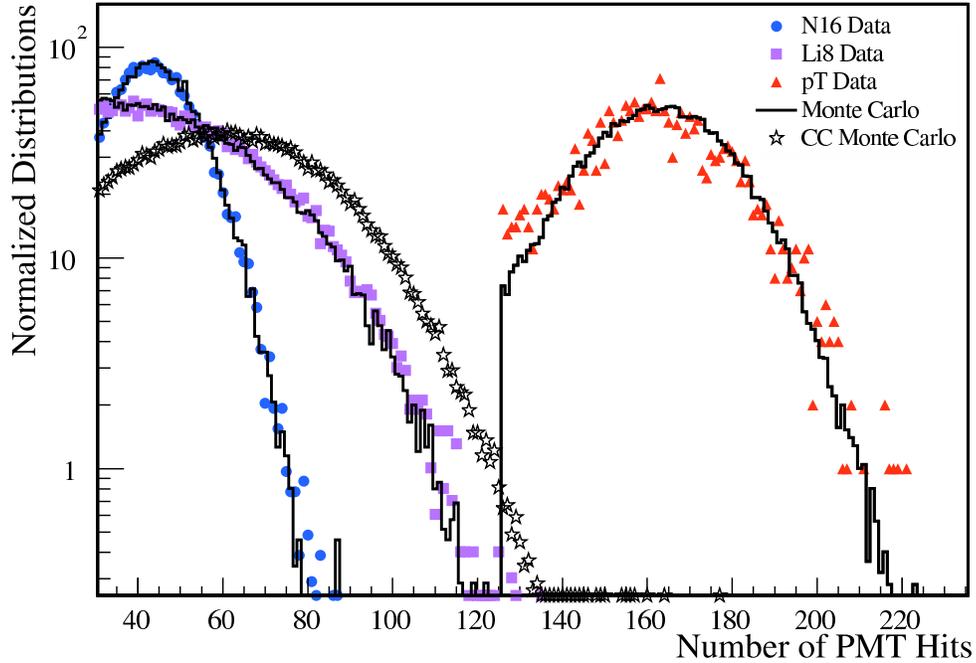}    
    \end{center}
    \caption{Comparison of the Monte Carlo predicted responses to 
    different calibrated sources.}
    \protect\label{fig:earmycc1}
\end{figure}

This tuned Monte Carlo is then used to make predictions for the 
energy response to different calibration sources.  The pT source 
generates 19.8-MeV $\gamma$ rays through the $^{3}$H(p,$\gamma$)$^{4}$He 
reaction~\cite{bib:poon}, and is used to check the linearity of the 
energy response beyond the endpoint of the $^{8}$B neutrino energy 
spectrum.  To probe the regions that are inaccessible to the calibration source deployment system, the $^{252}$Cf fission neutron source provides an extended 
distribution of 6.25-MeV $\gamma$ rays from d(n,$\gamma$)t.  
Figure~\ref{fig:earmycc1} shows a comparison of the Monte Carlo 
predictions and the detector responses to these sources.

The energy estimator uses the same input parameters (e.g. optical
parameters) as the Monte Carlo.  It assigns an effective kinetic
energy \teff\ to each event based upon its position, direction and the
number of hit PMTs within the prompt (unscattered) photon peak.  For 
an electron of total energy $E_{e}$, the derived 
detector energy response can be parameterized by a Gaussian:
\begin{displaymath}
    R(E_{\rm eff},E_{e})\;=\;\frac{1}{\sqrt{2\pi}\sigma_{E}(E_{e})}\exp
    \left[-\frac{1}{2}\left(\frac{E_{\rm eff}-E_{e}}{\sigma_{E}(E_{e})}\right)^{2}\right]
\end{displaymath}
where $E_{\rm eff}$=\teff+$m_{e}$, and the energy resolution is given by 
\begin{displaymath}
    \sigma_{E}(E_{e})=-0.0684+0.331\sqrt{E_{e}-m_e}+0.0425 (E_{e}-m_e)\;\mbox{MeV}.
\end{displaymath}
The systematic uncertainty on this absolute energy calibration is 
found to be $\pm$1.2\%, which results in neutrino flux uncertainties 
of $^{+4.3}_{-4.2}$\% for the CC channel and $^{+6.1}_{-6.2}$\% for the NC channel.  This is the most significant systematic uncertainty in the flux measurement.  Other energy related 
systematic uncertainties to the flux include the energy resolution 
and the energy scale linearity.  A summary of the systematic uncertainties can be found in Table~\ref{tbl:sys_err}.  

A second energy estimator using \nhits\ is employed for validation
purposes.  These two energy estimators give consistent results in the
neutrino flux measurement.

\subsection{Neutron Efficiency}

For the NC measurement of the solar neutrino  flux, the probability
that a neutron produced in the NC interaction will capture 
on deuterium and the detection efficiency of the 6.25-MeV $\gamma$ ray emitted following the neutron capture have to be determined.  The capture probability depends on the isotopic abundances of various nuclei in the
heavy water and the relevant capture cross sections for each.  Near the
boundary of the AV, the capture cross sections for neutrons in acrylic
and light water become important as well.  The detection efficiency of the 6.25-MeV $\gamma$ ray can be readily deduced from the \nsix\ calibration data, since the latter produces predominantly 6.13-MeV $\gamma$ rays.

Three different methods are used to calibrate the capture efficiency: a method which directly counts the number 
of neutrons measured with a $^{252}$Cf source and compares to source expectations, a purely analytic
calculation compared to Monte Carlo, and a multiplicity measurement which extracts the efficiency
statistically from the distribution of the number of detected neutrons per fission.

\begin{figure}
    \begin{center}
 \includegraphics[height=5.25in]{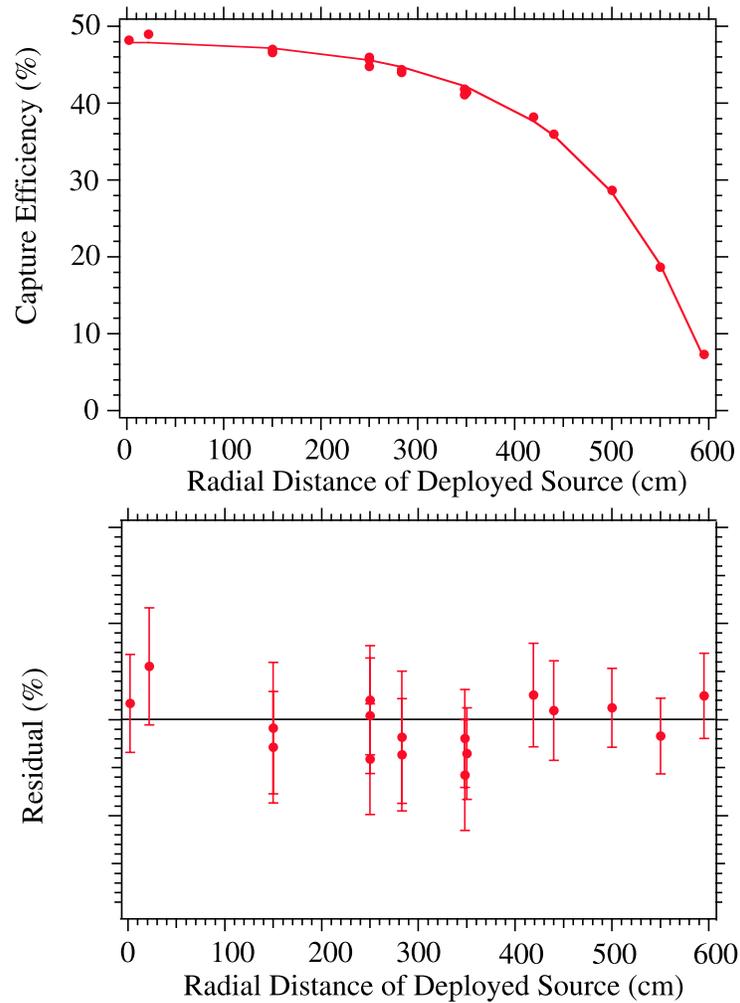}    
   \end{center}
    \caption{Final capture efficiencies measured with the direct counting method as a function of radial position, and the residuals from a fit to the radial dependence.}
    \protect\label{fig:NT-param-fit}
\end{figure}

The primary approach to measuring the capture efficiency is the
``direct counting" method, in which the number of neutrons detected during
a source run is compared to the total number expected to be generated by
the source based on the known decay rate.  This analysis was performed on $^{252}$Cf source data taken at various locations in the detector.  Because the $^{252}$Cf source is not a triggered source, non-neutron
backgrounds associated with the source and the detector were handled 
using reconstruction and instrumental background cuts, with a proper accounting of the associated
acceptance of these cuts in the efficiency measurement.    In Figure~\ref{fig:NT-param-fit} the neutron capture efficiency on deuterium as a function of radial distance from the center of the detector is shown.  The capture efficiency on deuterium for NC events is found to be $29.90\pm 1.10$\%, and when the 
fiducial volume cut of 550~cm and energy threshold cut of $T_{\rm eff} > 5.0$~MeV
are included, the overall detection efficiency is $14.38\pm0.53$\%. The 
uncertainties on these numbers contribute a relative uncertainty of
3.68\% in the extraction of the NC signal in the present analysis.  

The analytic calculation uses known cross sections and isotopic
abundances of the \dto\ to calculate the expected efficiency, and propagates uncertainties
on the microscopic parameters through the entire calculation.  The
largest uncertainties in the analytic model are on the fraction of \hto\ in
the heavy water, the capture cross sections on deuterium and $^{16}$O, and
the diffusion parameter which governs the total number of neutrons escaping
the \dto\ volume.  For the NC signal, the total uncertainty determined in
this way is $\sim$3\%.

The third approach is provided primarily as a check because of its
limited sensitivity.  In this approach the multiplicity distribution of
detected neutrons is fit with the expression
\begin{displaymath}
P(d) = \sum_{N=1}^{\infty} \sum_{r=d}^{\infty} \frac{r! \epsilon^d \epsilon^{r-d}}{d!(r-d)!}\frac{\exp^{\frac{-(r-N\mu)^2}{2N\sigma^2}}}{2N\pi \sigma^2}\frac{\lambda^N \exp^{-\lambda}}{N!}
\end{displaymath}
where $P(d)$ is the probability to detect $d$ neutrons per unit time given that
$r$ neutrons were generated according to Poisson statistics for the
$^{252}$Cf fission rate $\lambda$ and the known $^{252}$Cf multiplicity per
fission (mean $\mu = 3.79\pm 0.006$ and width $\sigma^2 = 1.57\pm 0.02$). In
this case, both the source fission rate $\lambda$ and the neutron detection 
efficiency $\epsilon$ are derived from a maximum likelihood fit.   The results from this approach and the direct counting approach are consistent.

\subsection{Low Energy Backgrounds}

Low levels of $^{232}$Th and $^{238}$U can be found naturally in all materials.  The SNO detector has been designed with very stringent radiopurity targets for different components in the detector.  Table~\ref{tbl:radio_target} lists the $^{232}$Th and $^{238}$U  target levels for the \dto, acrylic vessel (AV), and \hto.  At these radiopurity levels, the background to the NC signal is approximately 10\% of the NC signal predicted by the Standard Solar Model. 

\begin{table}
\begin{center}
\begin{tabular}{|c|c|c|} \hline
Component & $^{232}$Th  &  $^{238}$U \\ 
                      &  (g/g)   & (g/g)   \\ \hline
\dto  &  3.7$\times$10$^{-15}$  & 4.5$\times$10$^{-14}$ \\ \hline
\hto & 3.7$\times$10$^{-14}$  & 4.5$\times$10$^{-13}$ \\ \hline
AV & 1.9$\times$10$^{-12}$  & 3.6$\times$10$^{-12}$ \\ \hline
\end{tabular}
\end{center}
\caption{The target radio-purity levels for different components in the SNO detector.}
\protect\label{tbl:radio_target}
\end{table}

Radioactive decays of the daughters in the natural $^{232}$Th and $^{238}$U chains are the dominant backgrounds in the neutrino signal window.  These decays can contribute to the backgrounds in two different ways.  

A $\gamma$ ray with energy in excess of the binding energy of the deuteron (=2.2~MeV) can potentially photodisintegrate a deuteron:
\begin{displaymath}
	\gamma\,+ d \;\rightarrow\; n \, + p
\end{displaymath}
The free neutron in the final state is indistinguishable from that in the NC signature.  In the natural $^{232}$Th and the $^{238}$U chains, $\gamma$ rays that can photodisintegrate the deuterons are emitted in the  decays of \tltze\ and \bitof.  To measure the contribution of this photodisintegration background to the NC signal, it is necessary to determine the concentration of the different isotopes in the detector.  ``Ex-situ" radiochemical assays and "in-situ" Cherenkov light pattern recognition techniques are used to measure the contribution from this photodisintegration background.

A small fraction of decays with large $Q$ values (e.g. \tltze\ from the $^{232}$Th chain and \bitof\ from the $^{238}$U chain) inside the fiducial volume might generate enough light in the detector and register a calibrated energy above the analysis threshold.  Decays that are originated from outside the fiducial volume might get reconstructed to within the fiducial volume due to finite reconstruction resolution.  Low energy radioactive sources and simulations are used to understand this class of {\em Cherenkov tail} backgrounds.

\begin{figure}
    \begin{center}
     \includegraphics[width=5.25in]{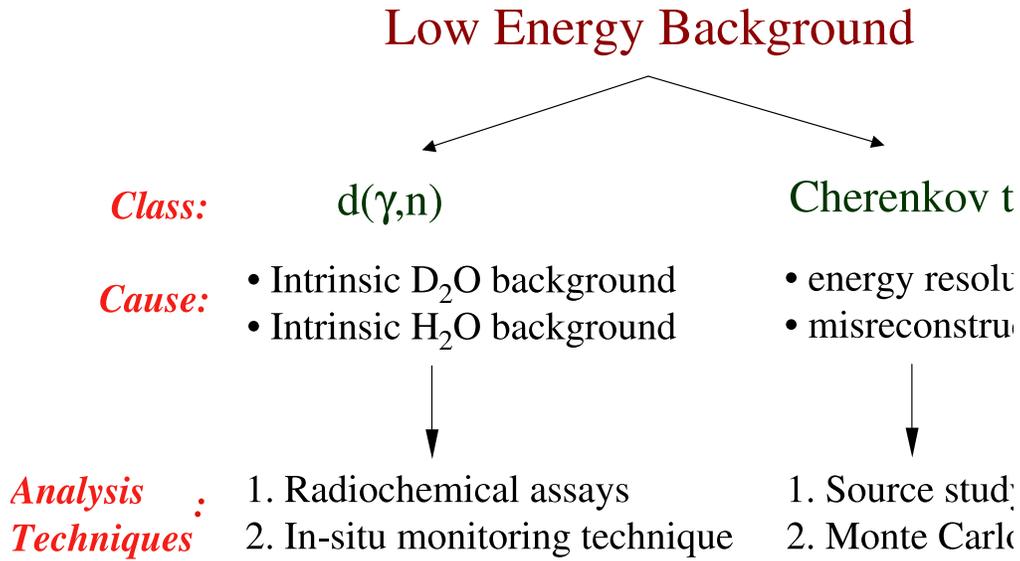}
   \end{center}
    \caption[Low Energy Background Analysis]{Low Energy Background Analysis.  The techniques that are used in understanding these backgrounds are shown.}
    \protect\label{fig:le_details}
\end{figure}

Figure~\ref{fig:le_details} shows the general strategy of the low energy background analysis.  In the following, we shall discuss the analysis techniques mentioned above in more details.

\subsubsection{Photodisintegration Background}

\vskip 14pt
\noindent {\bf \dto\ and \hto\ Radioactivity}\\

\begin{figure}
\begin{center}
 \includegraphics[height=3.25in]{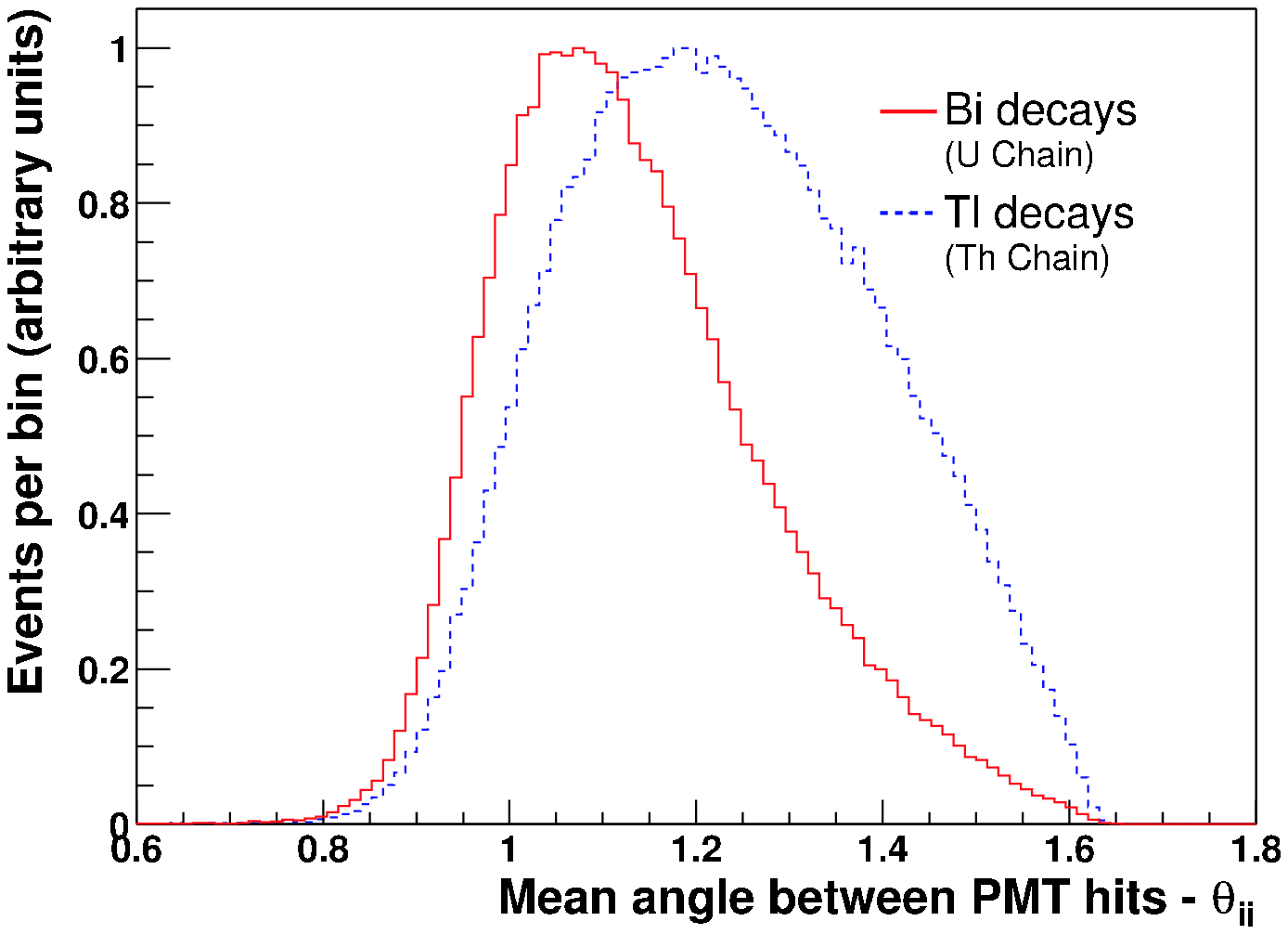}
    \end{center}
\caption{Difference in \tij\ between \tltze\ and \bitof\ decays.}
\protect\label{fig:thetaij}
\end{figure}

\begin{figure}
\begin{center}
 \includegraphics[width=4.5in]{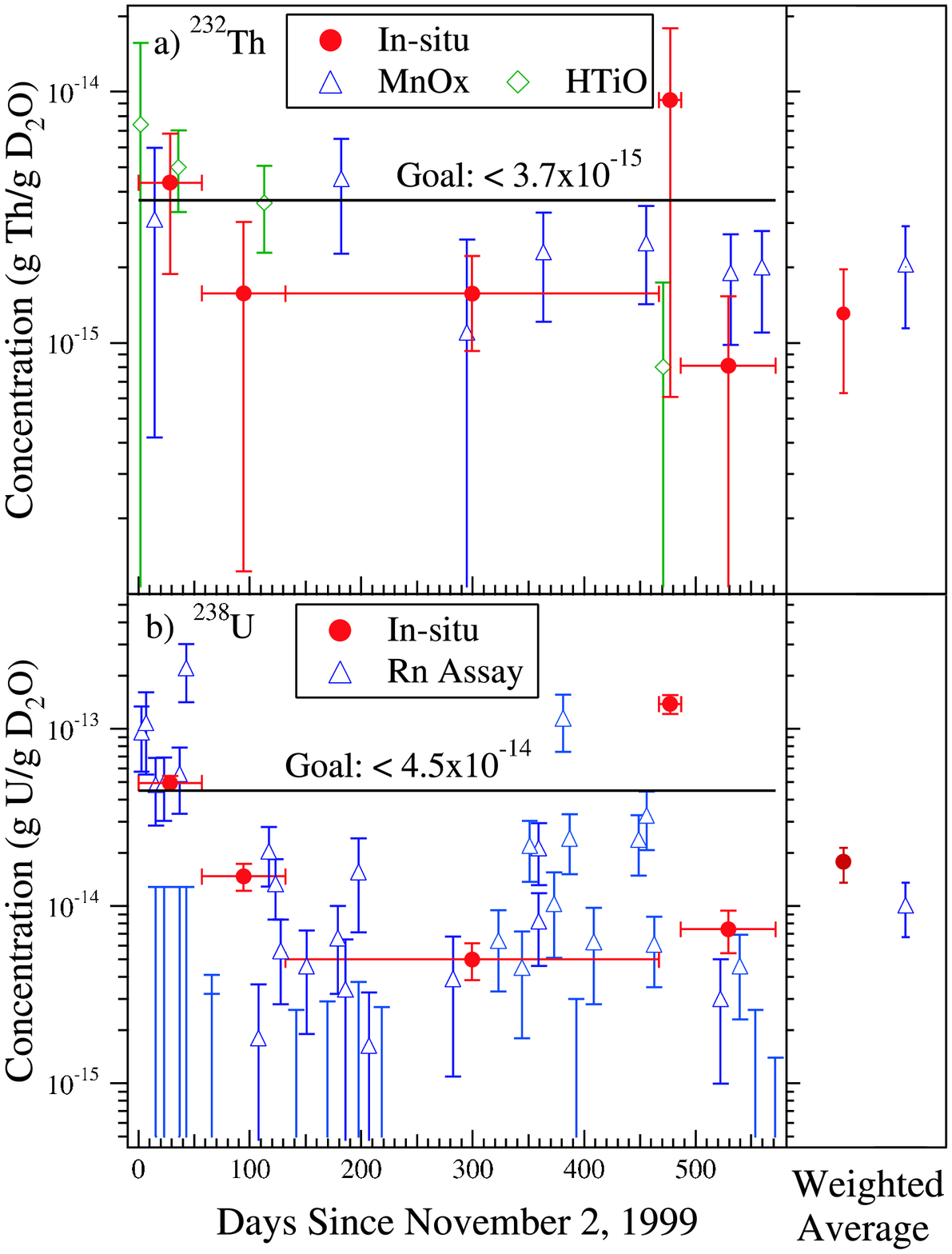}
    \end{center}
\caption{Thorium (a) and uranium (b) backgrounds (equivalent equilibrium concentrations) in the D$_2$O deduced by {\textit{in situ}} and {\textit{ex situ}} techniques.  The MnO$_{\rm{x}}$ and HTiO radiochemical assay results, the Rn assay results,  and the {\textit{in situ}} Cherenkov signal determination of the backgrounds  are presented for the period of this analysis on the left-hand side of frames (a) and (b).  The right-hand side shows time-integrated averages including an additional sampling systematic uncertainty for the {\textit{ex situ}} measurement. }
\protect\label{fig:pdback}
\end{figure}

Several techniques were developed to measure the $^{232}$Th and $^{238}$U concentration in the \dto\ and the \hto.  These techniques sample different daughters in the $^{232}$Th and $^{238}$U chains, and can be broadly classified into two categories:

\begin{itemize}
\item {\bf Ex-situ:}  These are techniques that involve chemical removal of radioisotopes from the \dto\ and the \hto, and their decays are counted in a system external to the SNO detector.  These techniques include the extraction of Ra isotopes using MnO$_{\mbox{x}}$ beads (the ``MnO$_{\mbox{x}}$" technique~\cite{bib:mnox_nim}), and the extraction of  Ra, Th and Pb isotopes using HTiO membranes (the ``HTiO" technique~\cite{bib:htio_nim}).  Typical assays circulated approximately 400 tonnes
of water through the extraction media.  These techniques provide isotopic identification of the decay daughters and contamination levels in the assayed water volumes, presented in Fig.~\ref{fig:pdback}~(a).   Secular equilibrium in the $^{238}$U decay chain is broken by the ingress of long-lived (3.8 day half-life) ${}^{222}$Rn in the experiment.  Measurements of this background are made by periodically extracting and cryogenically concentrating ${}^{222}$Rn from the $^{238}$U chain (the ``Rn assay"~\cite{bib:rn_nim}).  The Rn results are presented (as mass fractions in g(U)/g(D$_2$O)) in Fig.~\ref{fig:pdback}(b).

\vskip 14pt
\noindent {\em The MnO$_{\mbox{x}}$ Technique} \\

In the MnO$_{\mbox{x}}$ technique, \dto\ or \hto\ is passed through polypropylene columns that contain beads coated with a manganese oxide compound (MnO$_{\mbox{x}}$), which extracts Ra from the flowing water.  After a large volume of water has passed through the columns, they are removed and dried.  The dried column is then attached to a gas flow loop on an electrostatic counter (ESC).  The Rn produced from Ra decay is swept from the columns into the ESC where it decays.  The charged Po ions from the decay of Rn are carried by the electric field onto an $\alpha$ counter where the decays of the Po are detected, and their $\alpha$ energy spectra are collected.  For the $^{232}$Th chain, the relevant Po $\alpha$ decays are $^{216}$Po (6.8~MeV $\alpha$) and $^{212}$Po (8.8~MeV $\alpha$), whereas the relevant ones for the U chain are $^{218}$Po (6.0~MeV $\alpha$) and $^{214}$Po (7.7~MeV $\alpha$).  

\vskip 14pt
\noindent {\em The HTiO Technique}\\

In this technique, \dto\ or \hto\ is passed through hydrous titanium oxide (HTiO) trapped on filtration fibers.   The HTiO ion-exchanger is first deposited onto a microfiltration membrane.  Then the columns containing the loaded filters are used to extract $^{224}$Ra from the Th chain and $^{226}$Ra from the U chain from a large volume of \dto\ or \hto.  After extraction, Ra is eluted with nitric acid, and subsequently concentrated down to $\sim$10~m$l$ of eluate.  This is then mixed with liquid scintillator and counted using $\beta-\alpha$ delayed coincidence counters~\cite{bib:taplin_thesis}.  For the $^{232}$Th chain, the coincidences of the $\beta$-decay of $^{212}$Bi and the $\alpha$-decay of $^{212}$Po are counted, whereas the coincidences of the $\beta$-decay of $^{214}$Bi and the $\alpha$-decay of $^{214}$Po are counted for the $^{238}$U chain.

\vskip 14pt
\noindent {\em Rn Assays}\\

Measurements of $^{226}$Ra concentration in the \dto\ and the \hto\ in the MnO$_{\mbox{x}}$ and the HTiO techniques are not sufficient to determine the total radioactive background from the $^{238}$U chain.  This is because even a small leak of the underground laboratory air ($\sim$3~pCi/$l$ of $^{222}$Rn) can lead to significant disequilibrium between $^{226}$Ra and $^{214}$Bi.  The Rn assay technique was developed to tackle this problem.  In this technique, water drawn from discrete sample points in the detector is flowed through a degasser to liberate Rn.  The Rn is purified and collected in a cryogenic collector.  The subsequent $\alpha$ decays are counted in a Lucas cell scintillator (ZnS) chamber on a 2.54-cm diameter photomultiplier tube.  Since there is a delay of many $^{220}$Rn lifetimes between the preparation of the Lucas cells and their subsequent counting, this method is sensitive only to $^{222}$Rn decays.  

\item {\bf In-situ:}  This is a technique that uses pattern recognition on the Cherenkov light distribution to determine the concentration of  $^{232}$Th and $^{238}$U  in the \dto\ and the \hto~\cite{bib:chen_thesis,bib:mcgregor_thesis}.  The \tltze\ decay (from the $^{232}$Th chain) has a $Q$ value of $\sim$4.9~MeV, and the \bitof\ decay (from the $^{238}$U chain) has a $Q$ value of 3.27~MeV.  Almost every \tltze\ decay emits a 2.614~MeV $\gamma$, several low energy $\gamma$'s and a $\beta$ with an endpoint of $\sim$1-1.8~MeV, whereas there is a unique branch in the \bitof\ decay that produces a single $\beta$ with an endpoint energy of 3.27~MeV.  A single particle produces more Cherenkov photons than multiple particles with the same total energy because of the Cherenkov kinetic threshold.  Also the light isotropy is different for these two cases because multiple particles generate a more isotropic light pattern (multiple Cherenkov cones), whereas a single particle gives a more directional light distribution due to light emission within a single Cherenkov cone.  Therefore, by selecting a set of energy and fiducial volume cuts, it is possible to separate  \tltze\  and  \bitof\ decays statistically by their differences in light isotropy.  The light isotropy parameter that has been developed ($\langle$\tij$\rangle$) is the average angle between all hit PMT pairs within the prompt light time window in an event.  Figure~\ref{fig:thetaij} shows the difference in \tij\ between \tltze\ and \bitof\ decays.  In this in-situ analysis, Cherenkov events fitted within 450 cm from the detector center and extracted from the neutrino data set provide a time-integrated measure of these backgrounds over the same time period of the neutrino analysis.  Statistical separation of the \tltze\ and \bitof\ events is obtained by a maximum likelihood fit of the Cherenkov signal isotropy.

\end{itemize}

Results from the {\textit{ex situ}} and {\textit{in situ}} methods are consistent with each other as shown on the right hand side of Figs.~\ref{fig:pdback}(a)~and~\ref{fig:pdback}(b).
For the $^{232}$Th chain, the weighted mean (including additional sampling systematic uncertainty) of the two determinations is used for the background analysis.  The
$^{238}$U chain activity is dominated by Rn ingress which is highly time dependent.  Therefore the {\textit{in situ}} determination was used for this activity as it provides the appropriate time weighting.  For the present data set, the time-averaged equivalent equilibrium  $^{238}$U and $^{232}$Th concentration in the \dto\ are found to be
\begin{eqnarray}
^{232}\mbox{Th} &:& 1.63 \pm 0.58\times 10^{-15} \mbox{g Th / g \dto} \nonumber \\
^{238}\mbox{U} &:& 17.8 ^{+3.5}_{-4.3} \times 10^{-15} \mbox{ g U / g \dto}.  \nonumber
\end{eqnarray} 
Once the $^{232}$Th and $^{238}$U concentrations are measured, the photodisintegration background can be determined through Monte Carlo calculations that use the photodisintegration cross section, properties of $\gamma$-ray and neutron transport, and the  detector response function as inputs.  The expected number of observed photodisintegration neutrons are 18.4$\pm$6.5~counts for the $^{232}$Th background and 25.9$^{+5.0}_{-6.3}$~counts for the $^{238}$U background in the pure \dto\ phase of the experiment.

Daughters in the $^{232}$Th and $^{238}$U chain in the \hto\ and the acrylic vessel (AV) can still produce a photodisintegration background.  This happens when the $\gamma$'s with an energy $>$2.2~MeV enter the \dto\ volume and photodisintegrate a deuteron.   \textit{Ex situ} assays and \textit{in situ}
pattern recognition techniques described above were used to measure the $^{232}$Th and $^{238}$U concentrations in the \hto.  The photodisintegration background inside the fiducial volume due to radioactivity in the \hto\ was then determined.  The time-averaged equivalent equilibrium  $^{238}$U and $^{232}$Th concentration in the \hto\ are found to be
\begin{eqnarray}
^{232}\mbox{Th} &:& 14.2 \pm 6.6\times 10^{-14} \mbox{g Th / g \hto}  \nonumber \\
^{238}\mbox{U} &:& 75.5\pm 33.0 \times 10^{-14} \mbox{ g U / g \hto}.  \nonumber 
\end{eqnarray}
At these concentrations, the number of observed photodisintegration neutrons in the fiducial volume are determined to be 5.6$^{+3.6}_{-2.2}$~counts for the $^{232}$Th chain and 5.6$^{+4.2}_{-2.9}$~counts for the $^{238}$U chain.

\vskip 14pt
\noindent {\bf Acrylic Vessel Radioactivity}\\

Prior to the construction of the acrylic vessel, neutron activation analyses were done on samples of the acrylic panels used in the vessel.  These results, along with that of a dust density measurement on the vessel subsequent to its construction, indicate that the radioactive load of the acrylic vessel to be 7.5$\pm^{+1.7}_{-1.3}$~$\mu$g of $^{232}$Th and  15$\pm$15~$\mu$g of $^{238}$U.  Studies of the Cherenkov data revealed a radioactive hotspot on the acrylic vessel.   The radioactivity of this hotspot is equivalent to 10$^{+8.6}_{-3.6}$~$\mu$g of $^{232}$Th.  Combining the results from radioassays and the hotspot analysis, the expected number of observed photodisintegration background events due to radioactivity in the acrylic vessel is 15.8$^{+6.0}_{-6.8}$~counts.

Table~\ref{tbl:pd_summary} is a summary of the photodisintegration background for the data set used in this analysis.  In this table, we have also listed the contributions from other sources of neutron backgrounds.  Some of these are cosmic ray events and atmospheric neutrinos.  To reduce these backgrounds, an additional neutron background cut imposed a 250-ms veto (in software) following every event in which the total number of PMTs which registered a hit was greater than 60.   The contribution from these  additional sources are very small compared to the photodisintegration background.

\begin{table}
\begin{center}
\begin{tabular}{|c|c|c|c|} \hline 
               &   $^{232}$Th   &  $^{238}$U & {\bf Total}  \\
               &   (counts)    &   (counts) &  \\  \hline 
\dto\ photodisintegration    &  18.4$\pm$6.5 &   25.9$^{+5.0}_{-6.3}$ & 44.3$^{+8.2}_{-9.1}$ \\  \hline  
AV  photodisintegration    &   14.2$^{+5.8}_{-6.6}$ &  1.6$\pm$1.6  &  15.8$^{+6.0}_{-6.8}$ \\  \hline  
\hto\  photodisintegration   &  5.6$^{+3.6}_{-2.2}$ & 5.6$^{+4.2}_{-2.9}$& 11.2$^{+5.5}_{-3.6}$ \\  \hline  
Atmospheric $\nu$'s    and             &      \multicolumn{2}{c|}{}    &     \\ 
sub-Cherenkov threshold $\mu$'s       &   \multicolumn{2}{c|}{Not applicable} & $ 4 \pm 1$                   \\   \hline  
Spontaneous Fission                                 &  \multicolumn{2}{c|}{Not applicable} & $\ll1$              \\   \hline  
${}^{2}$H($\alpha,\alpha$)pn            & \multicolumn{2}{c|}{Not applicable} &  $2 \pm 0.4$         \\   \hline  
${}^{17}$O($\alpha$,n)                  &  \multicolumn{2}{c|}{Not applicable} & $\ll1$                       \\   \hline  
Terrestrial and reactor $\bar{\nu}$'s   & \multicolumn{2}{c|}{Not applicable} &  $1^{+3}_{-1}$         \\   \hline  
External neutrons                       &  \multicolumn{2}{c|}{Not applicable} & $\ll1$                       \\    \hline  
{\bf Total}  & \multicolumn{2}{c|}{} & $78 \pm 12$                 \\  \hline  
\end{tabular}
\end{center}
\caption{Summary of neutron background ($E>$5.5~MeV) in the fiducial volume ($R<$550~cm.) for the pure D$_2$O phase of the experiment}
\protect\label{tbl:pd_summary}
\end{table}

\subsubsection{Cherenkov Tail Background}

Low energy backgrounds from Cherenkov events in the neutrino signal region were evaluated by using 
acrylic encapsulated sources  of U and Th deployed throughout the detector volume
and by Monte Carlo calculations.    In the following, we discuss how this class of background is determined for different origins of radioactivity.

\vskip 14pt
\noindent {\small\bf Cherenkov Tail Background from \dto\ Radioactivity} \\

The Cherenkov tail background arising from internal radioactivity in the \dto\ were determined by a Monte Carlo study.  A Monte Carlo study is justified because the detector response in the \dto\ is well calibrated, and the internal radioactivity is well measured by the \textit{ex situ} and \textit{in situ} techniques.  In this Monte Carlo calculation, the number of Cherenkov background events is normalized to the number of photodisintegration (pd) neutrons  in the neutrino signal window and are found to be:
\begin{center}
\begin{tabular}{ll}
     $^{208}$Tl:  &  0.162  $^{+0.092}_{-0.030}$  \mbox{Cherenkov tail event/pd neutron event} \\
     $^{214}$Bi:  & 0.670  $^{+0.460}_{-0.125}$ \mbox{Cherenkov tail event/pd neutron event}.
\end{tabular}
\end{center}
Given the number of detected photodisintegration neutrons (Th: 18.4$\pm$6.5, U: 25.9$^{+5.0}_{-6.3}$), the expected Cherenkov tail backgrounds are:
\begin{center}
\begin{tabular}{ll}
     $^{232}$Th:  &  3.0 $^{+2.0}_{-1.3}$~counts  \\
     $^{238}$U:  & 17.4  $^{+12.4}_{-5.3}$~counts \\
    Total: & 20.4 $^{+12.6}_{-5.5}$~counts.
\end{tabular}
\end{center}

\begin{figure}
    \begin{center}
 \includegraphics[height=3.25in]{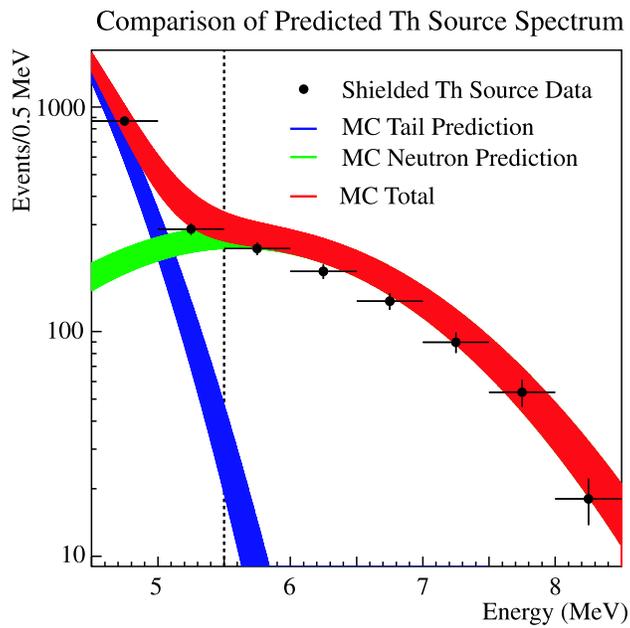}
    \end{center}
    \caption{Comparison of Monte Carlo prediction with systematic uncertainties to a stainless-steel encapsulated Th source at the center of the \dto.  The $\beta$'s in the decays are blocked by the stainless steel shroud, while the $\gamma$'s emitted in the decays escaped the stainless steel container.  The predictions, shown here as a $\pm$1$\sigma$ band, have been normalized by the livetime and the source strength.}
    \protect\label{fig:d2otail_src}
\end{figure}

As a consistency check, this Monte Carlo technique was used to make predictions on the energy spectrum of a variety of background sources, including a stainless-steel encapsulated Th source and a Rn spike in the detector during a water circulation pump failure.  Figure~\ref{fig:d2otail_src} compares the predicted energy spectrum for the stainless-steel encapsulated Th source run to the actual calibration source data.

\vskip 14pt
\noindent{\small\bf Cherenkov Tail Background from External Radioactivity}\\ 

In principle, one can perform Monte Carlo calculations to predict the Cherenkov tail background in the fiducial volume for radioactivity from the acrylic vessel, the \hto\ and the PMT support geodesic.  However, most of the events in the neutrino signal window with origins in the regions external to the fiducial volume come from the tail of the energy response and the reconstruction response functions.  These response functions are not as well understood outside the fiducial volume.  

\begin{figure}
    \begin{center}
 \includegraphics[height=4.25in]{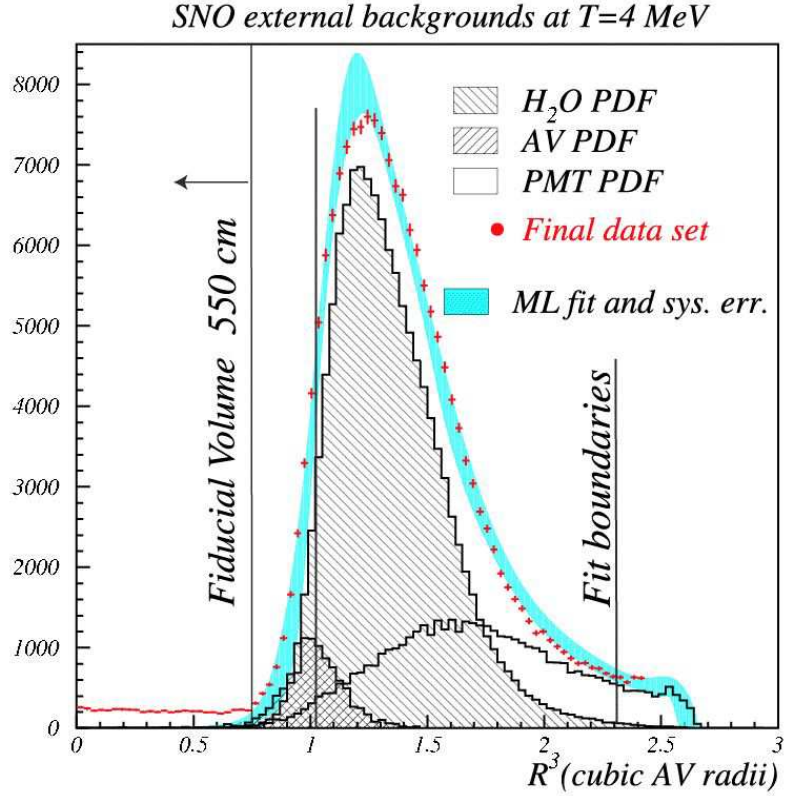}
    \end{center}
    \caption{External Cherenkov tail \rtree\ fit at \teff$>$4~MeV.  The extended maximum likelihood method was used in the fit, and the band represents the systematic uncertainties.}
    \protect\label{fig:external_ctail}
\end{figure}

A technique was developed to determine this external (acrylic vessel, \hto\ and PMT support geodesic) Cherenkov tail backgrounds in the neutrino signal window using the low energy background radial distribution ($R$) calibrated by low energy acrylic encapsulated sources.  The principle of this technique is to use the acrylic source data to generate the \rtree\ probability distribution functions (PDFs) for the low energy backgrounds in the AV, \hto\ and the PMT regions.  The \rtree\ distribution in these regions in the neutrino data was then fitted to a linear combination of these PDFs by the extended maximum likelihood method.  One difficulty of this technique is the lack of sufficient statistics at the intended neutrino analysis threshold (\teff=5~MeV).  In order to overcome this, the analysis was proceeded at a lower energy (\teff=4~MeV) and then extrapolated to 5~MeV.   The basic assumptions here are that there is no correlation between \rtree\ and energy, and that the reconstruction does not get worse with higher energy.   The \rtree\ fit was performed for the range $1.02<(R/600)^3<2.31$.  Figure~\ref{fig:external_ctail} shows the results of this \rtree\ fit.  The band shown in this plot is the range of the systematic uncertainties.  The data points are not centered in this band because the systematic uncertainties are not normally distributed.  Table~\ref{tbl:pd_summary} is a summary of the Cherenkov tail background (\teff$>$5~MeV) in the fiducial volume ($R<$550~cm.) for the pure D$_2$O phase of the experiment.  

\begin{table}
\begin{center}
\begin{tabular}{|c|c|c|c|} \hline 
               &   $^{232}$Th   &  $^{238}$U & {\bf Total}  \\
               &   (counts)    &   (counts) &  \\  \hline 
\dto    &   3.0$^{+2.0}_{-1.3}$ &   17.4$^{+12.4}_{-5.3}$& 20.4$^{+12.6}_{-5.5}$ \\   \hline  
AV     &   \multicolumn{2}{c|}{Not applicable} & 6.3$^{+2.9}_{-6.3}$  \\ \hline 
\hto    &  \multicolumn{2}{c|}{Not applicable} & 2.8$^{+3.9}_{-2.8}$ \\  \hline  
PMT  &  \multicolumn{2}{c|}{Not applicable} & 16.0$^{+11.1}_{-8.0}$ \\  \hline  
{\bf Total}  &  \multicolumn{2}{c|}{Not applicable}  &  {\bf 45.5$^{+17.5}_{-11.9}$}  \\ \hline  
\end{tabular}
\end{center}
\caption[Summary of Cherenkov tail background ($E>$5.5~MeV) in the fiducial volume ($R<$550~cm.) for the pure D$_2$O phase of the experiment]{Summary of Cherenkov tail background ($E>$5.5~MeV) in the fiducial volume ($R<$550~cm.) for the pure D$_2$O phase of the experiment.  The high energy $\gamma$ background contribution in the fiducial volume is included in the PMT $\beta\gamma$ entry.}
\protect\label{tbl:ctail_summary}
\end{table}

\begin{figure}
    \begin{center}
 \includegraphics[height=3.25in]{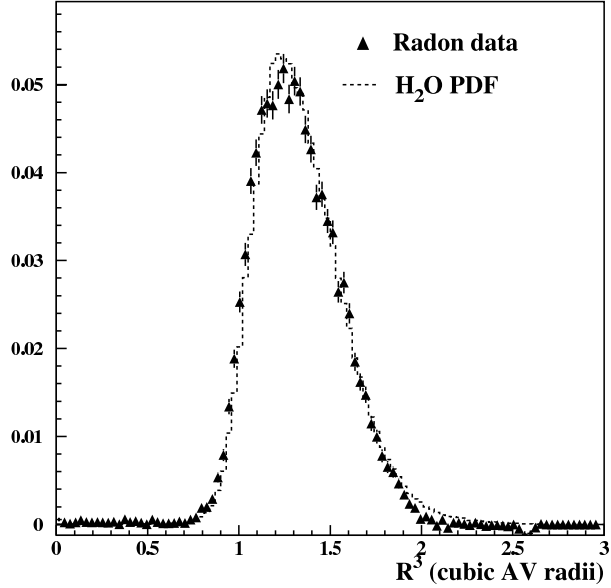}
    \end{center}
    \caption{Comparison between the PDF obtained from the radon spike in the detector and the \hto\ PDF derived from acrylic source data.}
    \protect\label{fig:hilopdfcomparison}
\end{figure}

Several consistency checks of this external Cherenkov tail analysis were performed.  The radial PDF for \hto\ background   was compared to the radial distribution of a radon spike during a water circulation pump failure.  This is shown in Figure~\ref{fig:hilopdfcomparison}.  The expected background rate in Table~\ref{tbl:ctail_summary} was also checked against the Monte Carlo calculations and the results of a simultaneous fit of $\nu$ signals and backgrounds in \nhits, \rtree\ and angular correlation to the Sun.  All these results are consistent.

\subsection{Signal Extraction}
\label{sec:sigex}

\begin{figure}
    \begin{center}
 \includegraphics[height=6in]{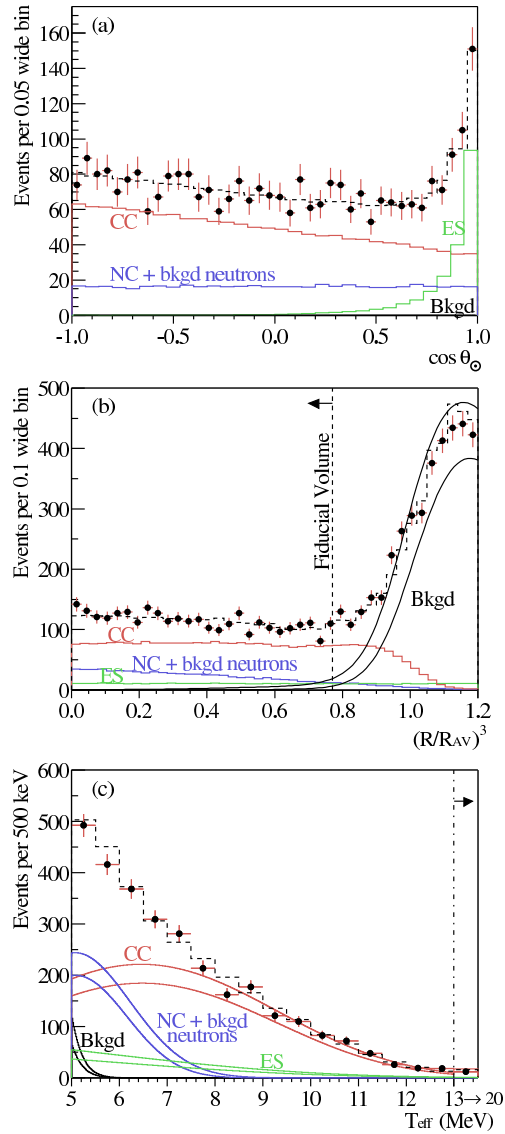}
    \end{center}
    \caption{(a) Distribution of \costs\ for $R \le 550$ cm. (b) Distribution of the volume weighted radial variable $(R/R_{\rm AV})^{3}$.  (c) Kinetic energy for $R \le 550$ cm.  Also shown are the Monte Carlo predictions for CC, ES and NC + background neutron events scaled to the fit results, and the calculated spectrum of Cherenkov background (Background) events.  The dashed lines represent the summed components, and the bands show $\pm 1\sigma$ uncertainties.  All distributions are for events with $T_{\rm eff}$$\geq$5 MeV. }
    \protect\label{fig:r3_comp}
\end{figure}

In order to test the null hypothesis, the assumption that there are only electron neutrinos in the solar neutrino flux,  the extended maximum likelihood method is used in extracting the CC,
ES and neutron (i.e. NC+background) contributions in the candidate data set.  Background contributions are constrained to the measured values discussed above.  The undistorted $^{8}$B spectrum from Ortiz {\em et al.}~\cite{bib:ortiz} is assumed in the signal decomposition.  Data
distributions in \teff,  the volume-weighted radial 
variable $(R/R_{AV})^{3}$ where $R_{\rm AV}=600$~cm is the radius of the acrylic vessel, and \costs\ are simultaneously fitted to the probability density functions (PDFs)
generated from Monte Carlo simulations.   \costs\ is the angle between the reconstructed direction of the event and the instantaneous direction from the Sun to
the Earth.  This distribution is shown in Figure~\ref{fig:r3_comp}(a) for the
analysis threshold of $T_{\rm eff}$$\geq 5$ MeV and fiducial volume selection
of $R\le 550$ cm, where $R$ is the reconstructed event radius.  The
forward peak (\costs$\sim$1) arises from the strong directionality in
the ES reaction.  The \costs\ distribution for the CC reaction, before
accounting for the detector response, is expected to be
(1-0.340\costs)~\cite{bib:ebgf}.  Fig.~\ref{fig:r3_comp}(b) shows the distribution of events in the volume-weighted radial 
variable $(R/R_{\rm AV})^3$, and Figure~\ref{fig:r3_comp}(c) shows the kinetic energy spectrum of the selected events.

The extraction yields $\nccfit$ CC events, $\nesfit$ ES events, and
$\nncfit$  NC events~\footnote{We note that this rate of neutron events also leads to a lower bound on the proton lifetime for ``invisible'' modes (based on the free neutron that would be left in deuterium\cite{bib:tretyak}) in excess of $10^{28}$ years, approximately 3 orders of magnitude more restrictive than previous limits\cite{bib:evans}. The possible contribution of this mechanism to the solar neutrino NC background is ignored.}. The uncertainties given above are statistical only, and the systematic uncertainties are summarized in Table~\ref{tbl:sys_err}.  The extracted counts for each neutrino detection channel above can be converted into integrated fluxes above the kinetic energy threshold of 5~MeV.  Assuming an undistorted $^{8}$B neutrino spectrum~\cite{bib:ortiz} and using only cross sections for \nue, the measured neutrino fluxes from each of the channels are (in units of $10^6~{\rm cm}^{-2} {\rm s}^{-1}$):
\begin{eqnarray*}
\phi_{\mbox{\tiny CC}} & = & \snoccfluxshort \\
\phi_{\mbox{\tiny ES}} & = & \snoesfluxshort \\
\phi_{\mbox{\tiny NC}} & = & \snoncfluxshort. 
\end{eqnarray*}
The CC and ES results reported here are consistent with our earlier results~\cite{bib:snocc} for $T_{\rm eff}$$\geq$6.75 MeV.  The excess of the NC flux over the CC and ES fluxes implies neutrino flavor transformations.

\begin{figure}
    \begin{center}
 \includegraphics[height=3.5in]{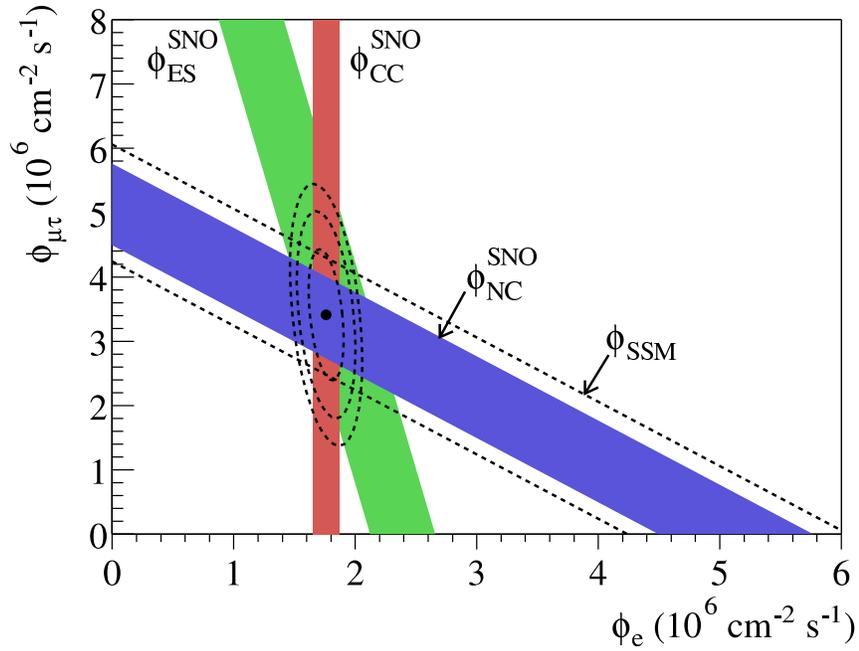}
    \end{center}
    \caption{Flux of ${}^{8}$B solar neutrinos which are $\mu$ or $\tau$ flavor {\it vs.} flux of electron neutrinos deduced from the three neutrino reactions in SNO.  The diagonal bands show the total ${}^{8}$B flux as predicted by the SSM~\cite{bib:bpb} (dashed lines) and that measured with the NC reaction in SNO (solid band).  The intercepts of these bands with the axes represent the $\pm 1\sigma$ errors.  The bands intersect at the fit values for $\phi_{e}$ and $\phi_{\mu\tau}$, indicating that the combined flux results are consistent with neutrino flavor transformation assuming no distortion in the ${}^{8}$B neutrino energy spectrum.}
    \protect\label{fig:phi_emutau}
\end{figure}

A direct test of the null hypothesis, i.e. there are only \nue\ in the solar neutrino flux, can be readily performed with a simple change of variables:
\begin{eqnarray*}
\phi_{\mbox{\tiny CC}} &=& \phi_e  \\
\phi_{\mbox{\tiny NC}} &=& \phi_e + \phi_{\mu\tau} \\
\phi_{\mbox{\tiny ES}} &=& \phi_e + \epsilon\phi_{\mu\tau},
\end{eqnarray*}
where \phie\ is the flux of the electron component, \phinumutau\ is the flux of non-electron component, and $\epsilon$=0.1559 is the ratio of the elastic scattering cross sections for $\nu_{\mu\tau}$ and $\nu_e$ above the kinetic threshold of 5~MeV.  Assuming an undistorted $^{8}$B energy spectrum, a maximum likelihood extraction using these transformed variables gives the fluxes of the electron and non-electron components as:
\begin{eqnarray*}
\phi_{e} & = & \snoeflux \\
\phi_{\mu\tau} & = & \snomutauflux 
\end{eqnarray*}
Combining the statistical and systematic uncertainties in quadrature,
 $\phi_{\mu\tau}$ is $\snomutaufluxcomb$,
which is \nsigmassno$\sigma$  above zero, and provides   strong
evidence for flavor transformation consistent with neutrino oscillations~\cite{bib:maki,bib:pontecorvo}.  Adding the Super-Kamiokande ES measurement of the ${}^{8}$B flux~\cite{bib:superk} $\phi^{\mbox{\tiny SK}}_{\mbox{\tiny ES}}=\phisk$ as an additional constraint, we find $\phi_{\mu\tau}=\snomutaufluxsk$, which is \nsigmassk$\sigma$ above zero.
Figure~\ref{fig:phi_emutau} shows the flux of non-electron flavor active neutrinos
{\it vs.} the flux of electron neutrinos deduced from the SNO data.  The three bands represent the one standard deviation measurements of the CC, ES, and NC rates. The error ellipses represent  the 68\%, 95\%, and 99\% joint probability contours for $\phi_{e}$ and $\phi_{\mu\tau}$.

Removing the constraint that the solar neutrino energy spectrum is undistorted, the signal decomposition is repeated using only the $\cos \theta_{\odot}$ and $(R/R_{\rm AV})^3$ information.  The total flux of active ${}^{8}$B neutrinos measured with the NC reaction is 
\begin{displaymath}
\phi_{\mbox{\tiny NC}}=  \snoncfluxunc 
\end{displaymath}
\noindent which is in agreement with the shape constrained value above and with the standard solar model prediction~\cite{bib:bpb} for ${}^{8}$B $\nu$'s, $\phi_{\mbox{\tiny SSM}} = \ssmflux.$ 

\begin{table}
\begin{tabular}{llllll}\hline
Source        &  CC Uncert.          & NC Uncert.         & $\phi_{\mu\tau}$ Uncert.    \\
                    & (percent)         &(percent)          &(percent)                  \\ \hline
Energy scale $\dagger$  & -4.2,+4.3 &-6.2,+6.1 &-10.4,+10.3  \\ 
Energy resolution $\dagger$  & -0.9,+0.0 &-0.0,+4.4 &-0.0,+6.8  \\ 
Energy non-linearity $\dagger$  & $\pm 0.1$ &  $\pm 0.4$ &  $\pm 0.6$    \\ 
Vertex resolution $\dagger$  & $\pm 0.0$ &  $\pm 0.1$ &  $\pm 0.2$    \\ 
Vertex accuracy & -2.8,+2.9 &$\pm 1.8$ &  $\pm 1.4$    \\ 
Angular resolution & -0.2,+0.2  &  -0.3,+0.3 &  -0.3,+0.3 \\ 
Internal source pd $\dagger$  & $\pm 0.0$ &  -1.5,+1.6 &-2.0,+2.2  \\ 
External source pd & $\pm 0.1$ &  -1.0,+1.0 &$\pm 1.4$    \\ 
D$_2$O Cherenkov $\dagger$  & -0.1,+0.2 &-2.6,+1.2 &-3.7,+1.7  \\ 
H$_2$O Cherenkov & $\pm 0.0$ &  -0.2,+0.4 &-0.2,+0.6  \\ 
AV Cherenkov & $\pm 0.0$ &  -0.2,+0.2 &-0.3,+0.3  \\ 
PMT Cherenkov $\dagger$  & $\pm 0.1$ &  -2.1,+1.6 &-3.0,+2.2  \\ 
Neutron capture & $\pm 0.0$ &  -4.0,+3.6 &-5.8,+5.2  \\ 
Cut acceptance & -0.2,+0.4 &-0.2,+0.4 &-0.2,+0.4  \\ \hline 
Experimental uncertainty & -5.2,+5.2 &  -8.5,+9.1 &  -13.2,+14.1 \\ \hline 
Cross section~\cite{bib:crosssection} & $\pm 1.8$  & $\pm 1.3 $& $\pm 1.4$ \\ \hline
\end{tabular}
\caption{Systematic uncertainties on fluxes. The experimental uncertainty for ES (not shown) is -4.8,+5.0 percent. $\dagger$ denotes CC {\it vs.} NC anti-correlation.}
\protect\label{tbl:sys_err}
\end{table}

\subsection{Day-Night Analysis}

The solar neutrino flux results presented above provide strong evidence for neutrino flavor transformation independent of solar model predictions.  One of the possible mechanism for this flavor transformation is mass-induced neutrino oscillations.  For some values of the
mixing parameters, spectral distortions and a measurable dependence on
solar zenith angle are expected~\cite{bib:theo1,bib:theo2,bib:theo3}.
This solar zenith angle dependence might be caused by interaction with the matter of the Earth
(the MSW effect) and would depend not only on oscillation parameters
and neutrino energy, but also on the neutrino path length and $e^-$ density
through the Earth. 

To look for this distinctive signature of neutrino oscillation, a solar neutrino flux analysis similar to that described above was performed for two solar zenith angle $\theta_z$ bins:  $\theta_z>0$ (``day") and $\theta_z<0$ (``night").    During the development of this analysis, the data
were partitioned into two sets of approximately equal livetime (split
at July 1, 2000), each having substantial day and night components.
Analysis procedures were refined during the analysis of Set 1 and
fixed before Set 2 was analyzed with the same procedures. The latter thus served as an unbiased
test.  Unless otherwise stated, the analysis presented in the following is for the combined data set.

\begin{figure}
\begin{center}
 \includegraphics[height=3.5in]{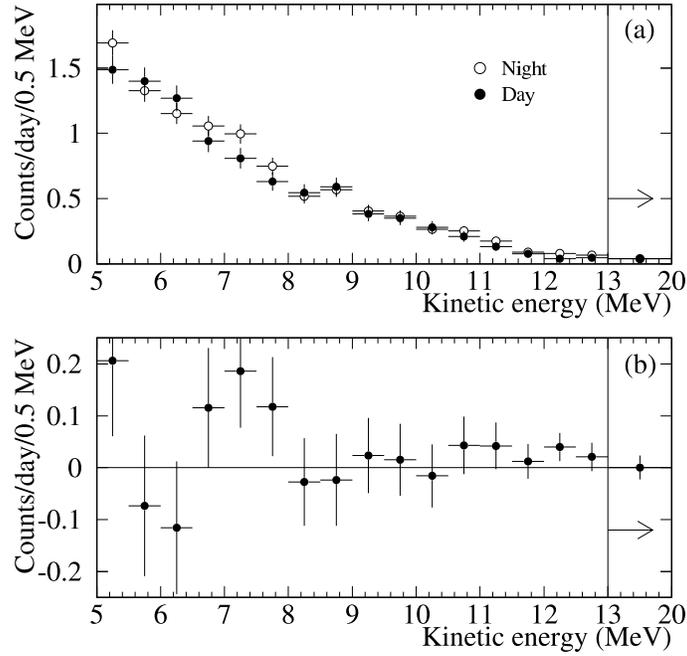}
    \end{center}
\caption{(a) Energy spectra (signals + background) for day and night bins.
The final energy bin extends from 13.0 to 20.0 MeV.  (b) Difference between the spectra (night - day).  The day rate was $9.23 \pm 0.27$ events/day, and the night rate was $9.79 \pm 0.24$ events/day.}
\protect\label{fig:prl_dn_spectra}
\end{figure}

For each neutrino interaction channel, the asymmetry ratio (${\mathcal{A}}$) of the measured day flux ($\phi_D$) and night flux  ($\phi_N$):
\begin{displaymath}
{\mathcal{A}}_i \;=\; 2\, \frac{\phi_N\,-\,\phi_D}{\phi_N\,+\,\phi_D} \hspace{0.5in} i=\mbox{CC,NC,ES}
\end{displaymath}
was determined.  Figure~\ref{fig:prl_dn_spectra} shows the day and night energy spectrum for all the accepted events above the kinetic energy threshold of 5~MeV and inside the fiducial volume of the inner 550~cm.   In the extraction of the neutrino fluxes, backgrounds were subtracted separately for the two zenith angle bins.  The results were then corrected for the orbital eccentricity by normalizing to an Earth-Sun distance of 1~AU.  Table~\ref{tbl:dn_sigex} is a summary of these extractions, where the day and night fluxes are given for
the combined data.  A $\chi^2$ consistency test of the six measured
fluxes between Sets 1 and 2 yielded a chance probability of 8\%.  A
similar test done directly on the three asymmetry parameters gave a
chance probability of 2\%.  No systematic has been identified, in
either signal or background regions, that would suggest that the
differences between Set 1 and Set 2 are other than a statistical
fluctuation.  For the combined analysis, ${\mathcal{A}}_{CC}$ is
$+2.2\sigma$ from zero, while ${\mathcal{A}}_{ES}$ and
${\mathcal{A}}_{NC}$ are $-0.9\sigma$ and $-1.2\sigma$ from zero,
respectively.  Note that ${\mathcal{A}}_{CC}$ and ${\mathcal{A}}_{NC}$ are
strongly statistically anti-correlated ($\rho=-0.518$), while
${\mathcal{A}}_{CC}$ and ${\mathcal{A}}_{ES}$ ($\rho=-0.161$) and
${\mathcal{A}}_{ES}$ and ${\mathcal{A}}_{NC}$ ($\rho=-0.106$) are
moderately anti-correlated.  Table \ref{tbl:dn_syserr} gives the
systematic uncertainties on the asymmetry parameters.

\scriptsize

\begin{table}
\begin{center}
\begin{tabular}{c|cc|c} \hline
 & 
\multicolumn{2}{c|}{Combined} &   ${\mathcal{A}} (\%)$ \\ 
signal & $\phi_{D}$ & $\phi_{N}$ & \\  
\hline
CC & 
$1.62\pm0.08\pm 0.08$  & $1.87\pm 0.07\pm 0.10$ & $+14.0 \pm~6.3
^{+1.5}_{-1.4}$ \\ 
ES & 
$2.64\pm0.37\pm 0.12$ & $2.22 \pm0.30\pm 0.12$ &  $-17.4 \pm 19.5
^{+2.4}_{-2.2}$ \\ 
NC & 
$5.69\pm0.66\pm 0.44$  & $4.63\pm0.57\pm 0.44$  &  $-20.4 \pm 16.9
^{+2.4}_{-2.5}$ \\  \hline 
\end{tabular}
    \end{center}
\caption{The results of signal extraction for the two zenith angle bins.  The fluxes shown here have been normalized to an Earth-Sun distance of 1 AU.  An undistorted  $^8$B spectrum was assumed in the signal decomposition.  The systematic uncertainties (combined set) include a component that cancels in the formation of the${\mathcal{A}}$.  Except for the dimensionless ${\mathcal{A}}$, the units are $10^6$~cm$^{-2}$~s$^{-1}$.  Flux values have been rounded, but the
asymmetries were calculated with full precision.}
\protect\label{tbl:dn_sigex} 
\end{table}

\normalsize

\begin{table}
\begin{center}
\begin{tabular}{c|c|c|c} \hline
Systematic & $\delta {\mathcal{A}}_{CC} $ & $\delta {\mathcal{A}}_{ES} $ &
$\delta {\mathcal{A}}_{NC} $ \\ 
\hline
Long-term energy scale drift       & 0.4 & 0.5 & 0.2 \\
Diurnal energy scale variation     & 1.2 & 0.7 & 1.6 \\
Directional energy scale var.      & 0.2 & 1.4 & 0.3 \\
Diurnal energy resolution var.     & 0.1 & 0.1 & 0.3 \\
Directional energy resolution var. & 0.0 & 0.1 & 0.0 \\
Diurnal vertex shift var.              & 0.5 & 0.6 & 0.7 \\
Directional vertex shift var. & 0.0 & 1.1 & 0.1 \\
Diurnal vertex resolution var.     & 0.2 & 0.7 & 0.5 \\
Directional angular recon. var.    & 0.0 & 0.1 & 0.1 \\
PMT $\beta$-$\gamma$ background    & 0.0 & 0.2 & 0.5 \\
AV+H$_2$O $\beta$-$\gamma$ background.          & 0.0 & 0.6 & 0.2 \\
D$_2$O $\beta$-$\gamma$, neutrons background.   & 0.1 & 0.4 & 1.2 \\
External neutrons background.                 & 0.0 & 0.2 & 0.4 \\
Cut acceptance                          & 0.5 & 0.5 & 0.5 \\
\hline
Total                              & 1.5 & 2.4 & 2.4 \\ \hline
\end{tabular}
    \end{center}
\caption{Effect of systematic uncertainties on
${\mathcal{A}}~(\%)$.  For presentation, uncertainties have been
symmetrized and rounded.}
\protect\label{tbl:dn_syserr}
\end{table}

Systematic checks were made on a set of signals that are continuously present in the detector in order to look for any diurnal variation in the detector response.  These studies include livetime verification using the detector diagnostic triggers (pulsed at 5~Hz),  variation in the observed muon rate, variation in the detector response to the muon-induced secondary neutrons, variation in the detector response to the radioactive hotspot on the acrylic vessel, and a neutrino signal extraction based on an east-west division instead of a zenith angle division.  These studies do not show any significant diurnal variation in the detector response.  In the study of the diurnal variation of the radioactive hotspot, a limit of 3.5\% on the rate asymmetry was determined.  Because of its steeply falling energy spectrum, a 0.3\% limit was set for the diurnal variation of the detector's energy scale.

\begin{figure}
\begin{center}
 \includegraphics[height=3.25in]{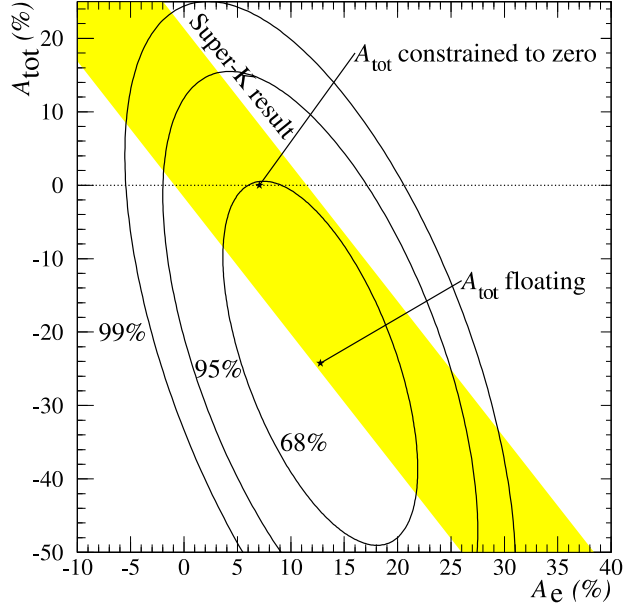}
    \end{center}
\caption{ Joint probability contours for
${\mathcal{A}}_{tot}$ and ${\mathcal{A}}_{e}$.  The points indicate the
results when ${\mathcal{A}}_{tot}$ is allowed to float and when it is
constrained to zero.  The diagonal band indicates the 68\% joint
contour for the Super-K ${\mathcal{A}}_{ES}$ measurement (${\mathcal{A}}_{ES}=3.3\% \pm
2.2\%~\mathrm{(stat.)}^{+1.3}_{-1.2}\%~\mathrm{(sys.)}$~\cite{bib:superk}. ) }
\protect\label{fig:dn_contour}
\end{figure}

The asymmetry ratio in the \nue\ flux ${\mathcal{A}}_{e}$ and the total active neutrino flux ${\mathcal{A}}_{total}$ can be readily determined by a change of variables in the signal extraction (see Sec.~\ref{sec:sigex}).  Figure~\ref{fig:dn_contour} shows the ${\mathcal{A}}_e$
vs. ${\mathcal{A}}_{tot}$ joint probability contours.  Forcing
${\mathcal{A}}_{tot} = 0$, as predicted by active-only models, yielded
the result of ${\mathcal{A}}_{e} = 7.0\%
\pm 4.9\%~\mathrm{(stat.)  ^{+1.3}_{-1.2}}\%~\mathrm{(sys.)}$.

\section{Analysis of Neutrino Mixing Parameters}

\begin{figure}
\begin{center}
 \includegraphics[width=3.25in]{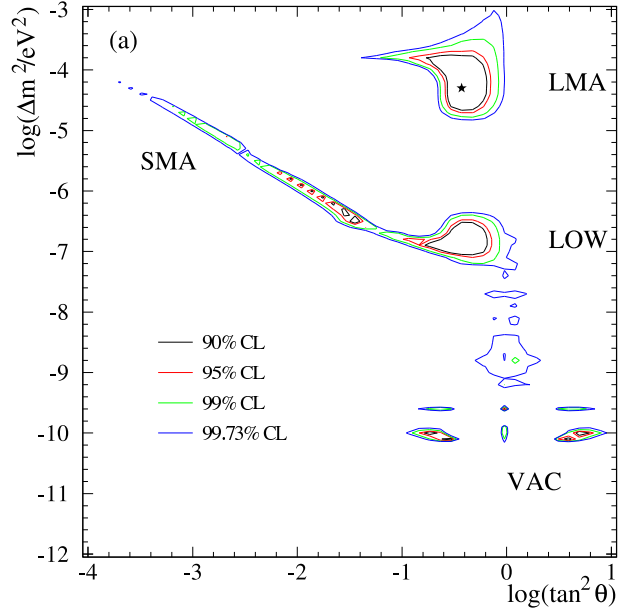}
    \end{center}
\caption{Allowed regions of the MSW plane determined  by a $\chi^2$ fit to 
SNO day and night energy spectra.  The star indicates the best fit.} 
\protect\label{fig:sno_only}
\end{figure}

Using the day and night energy spectrum in Figure~\ref{fig:prl_dn_spectra}, an analysis to constrain the allowed MSW mixing parameters under the 2-$\nu$ flavor mixing framework was performed.  The radial distribution $R$ and the solar angular correlation (\costs) information were omitted.   The expected number of counts were calculated for each of the energy bins by convoluting the $^8$B neutrino spectrum~\cite{bib:ortiz}, the MSW survival probability, the neutrino interaction cross sections and the energy response of the SNO detector.  With the high energy $hep$ neutrino flux fixed at $9.3 \times 10^{3}$~cm$^{-2}$~s$^{-1}$~\cite{bib:bpb}, the total $^8$B flux $\phi_{B}$, the difference $\Delta m^2$ between the squared masses of
the two neutrino mass eigenstates, and the mixing angle $\theta$ are the only free parameters in a $\chi^2$ fit of the MSW model to the observed data.  Figure~\ref{fig:sno_only} shows the allowed regions in the $\Delta m^2$-$\theta$ space at the 90\%, 95\%, 99\% and 99.73\% confidence levels.

\begin{figure}
\begin{center}
 \includegraphics[width=3.25in]{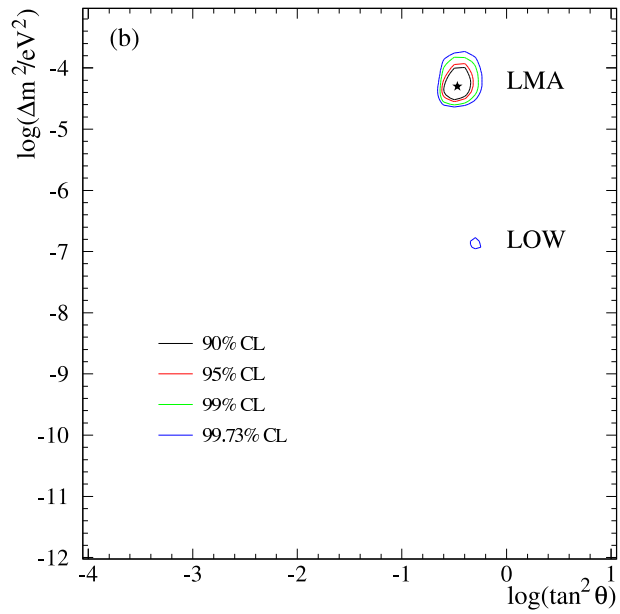}
    \end{center}
\caption{Allowed regions of the MSW plane determined  by a $\chi^2$ fit to 
SNO day and night energy spectra, the solar neutrino fluxes from the Cl and Ga experiments and the day and night energy spectra from the Super-Kamiokande experiment.  The star indicates the best fit.} 
\protect\label{fig:global}
\end{figure}

Additional information from other solar neutrino experiments can be used in constructing the allowed $\Delta m^2$-$\theta$ regions.  In Figure~\ref{fig:global}, the solar neutrino fluxes measured by the Cl experiment~\cite{bib:homestake} and the Ga experiments~\cite{bib:sage,bib:gallex,bib:gno}, along with the day and night neutrino energy spectra from the Super-Kamiokande experiment~\cite{bib:superk} are combined with the day and night energy spectra from SNO in a fit of the three free parameters ($\phi_{B}$, $\Delta m^2$ and $\theta$).  The $pp$, $pep$, $^7$Be and $hep$ neutrino fluxes were fixed at their predicted values in the Standard Solar Model~\cite{bib:bpb}.   Table~\ref{tbl:dn_chisq} summarizes the best fit points in the $\Delta m^2$-$\theta$ plane.   It is clear from this table and Figure~\ref{fig:global} that this global analysis strongly favors the Large Mixing Angle (LMA) region, and excludes the region of $\tan\theta>$1.

\begin{table}
\begin{center}
\begin{tabular}{c|cccccc}
Region & $\chi^2_{min}$/dof & $\phi_{B}$ & ${\mathcal{A}}_e (\%)$ &
$\Delta m^2$ & $\tan^2 \theta$ & c.l.(\%)\\
\hline
LMA &  57.0/72 & 5.86 & 6.4 & $5.0 \times 10^{-5}$ & 0.34 &  --- \\
LOW &  67.7/72 & 4.95 & 5.9 & $1.3 \times 10^{-7}$ & 0.55 &  99.5 \\
\end{tabular}
    \end{center}
\caption{Best fit points in the MSW plane for global
MSW analysis using all solar neutrino data.  $\phi_{B}$ is the
best-fit $^8$B flux for each point, and has units of
$10^6$~cm$^{-2}$~s$^{-1}$. $\Delta m^2$ has units of eV$^2$.
${\mathcal{A}}_e$ is the predicted asymmetry for each point.}
\protect\label{tbl:dn_chisq}
\end{table}

\section{Conclusions}

The results that are presented in this paper are truly groundbreaking.  The neutral-current measurement is the first direct measurement of the total active $^8$B neutrino flux.  The neutrino fluxes measured by this channel provide strong evidence for neutrino flavor transformation, thereby solving the long-standing Solar Neutrino Problem.  The significance of \nue\ transforming into $\nu_\mu$ or $\nu_\tau$ is at 5.3$\sigma$ level.  This neutral-current measurement also verified the Standard Solar Model prediction of the $^8$B neutrino flux.  

The analysis presented in this paper is also the first direct measurement of the day-night asymmetries in the \nue\ flux and the total neutrino flux.   When combining the day and  night energy spectra  from SNO with results from other solar neutrino experiments in a global 2-$\nu$ flavor analysis of the MSW oscillation parameters, the LMA solution is strongly favored and the ``dark side" ($\tan\theta>$1) is excluded.  

In the coming years, SNO will continue to make significant contributions to solar neutrino physics by refining the neutral-current measurement and the day-night asymmetry with different experimental techniques.  It will also attempt to look for other possible signatures of neutrino oscillation (e.g. spectral distortion) and other rare processes (e.g. nucleon decays, anti-$\nu$) in its future physics program.

\section{Acknowledgements}

This research was supported by the Natural Sciences and Engineering 
Research Council of Canada, Industry Canada, National Research 
Council of Canada, Northern Ontario Heritage Fund Corporation and the 
Province of Ontario, the United States Department of Energy and in the 
United Kingdom by the Science and Engineering Research Council and the 
Particle Physics and Astronomy Research Council.  Further support was 
provided by INCO, Ltd., Atomic Energy of Canada Limited (AECL), 
Agra-Monenco, Canatom, Canadian Microelectronics Corporation, AT\&T 
Microelectronics, Northern Telecom and British Nuclear Fuels, Ltd.  
The heavy water was loaned by AECL with the cooperation of Ontario 
Power Generation.  We thank the SNO technical staff for their strong contributions.

\end{document}